\documentclass[twocolumn]{aastex61}
\usepackage{booktabs}
\usepackage{hyperref}
\hypersetup{
    colorlinks=true,
    linkcolor=blue,
    filecolor=magenta,      
    urlcolor=cyan,
}
\usepackage{rotating}
\usepackage{multirow}
\usepackage{float}
\usepackage{bbding}
\usepackage{soul}

\def\arcsec{\ifmmode^{\prime\prime}\;\else$^{\prime\prime}\;$\fi}
\def\arcmin{\ifmmode^{\prime}}

\usepackage{color}

\definecolor{alizarin}{rgb}{0.82, 0.1, 0.26}
\definecolor{aquamarine}{rgb}{0.5, 1.0, 0.83}
\definecolor{brightpink}{rgb}{1.0, 0.0, 0.5}
\definecolor{ceruleanblue}{rgb}{0.16, 0.32, 0.75}
\definecolor{crimsonglory}{rgb}{0.75, 0.0, 0.2}
\definecolor{coral}{rgb}{1.0, 0.5, 0.31}

\received{??} 
\revised{??}
\accepted{??}
\submitjournal{ApJ}

\shorttitle{Infrared dust continuum in NGC\,1068}
\shortauthors{C. Victoria-Ceballos}

\begin{document}
\title{The complex infrared dust continuum emission of NGC\,1068: \\ ground-based N- and Q-band spectroscopy and new radiative transfer models}

\correspondingauthor{C. Victoria-Ceballos, PhD student}
\email{c.victoria@irya.unam.mx}

\author{C\'esar Ivan Victoria-Ceballos} 
\affil{Instituto de Radioastronom\'ia y Astrof\'isica (IRyA-UNAM), 3-72 (Xangari), 8701, Morelia, M\'exico}
\author{Omaira Gonz\'alez-Mart\'in}
\affil{Instituto de Radioastronom\'ia y Astrof\'isica (IRyA-UNAM), 3-72 (Xangari), 8701, Morelia, M\'exico}
\author{Jacopo Fritz}
\affil{Instituto de Radioastronom\'ia y Astrof\'isica (IRyA-UNAM), 3-72 (Xangari), 8701, Morelia, M\'exico}
\author{Cristina Ramos Almeida}
\affil{Instituto de Astrof\'isica de Canarias, Calle V\'ia L\'actea, s/n, E-38205 La laguna, Tenerife, Spain}
\affil{Departamento de Astrof\'isica, Universidad de La Laguna, E-38206 La Laguna, Tenerife, Spain}
\author{Enrique L\'opez-Rodr\'iguez}
\affil{Kavli Institute for Particle Astrophysics \& Cosmology (KIPAC), Stanford University, Stanford, CA 94305, USA}
\author{Santiago Garc\'ia-Burillo}
\affil{Observatorio Astron\'omico Nacional (OAN-IGN)-Observatorio de Madrid, Alfonso XII, 3, E-28014, Madrid, Spain}
\author{Almudena Alonso-Herrero}
\affil{Centro de Astrobiolog\'ia (CSIC-INTA), ESAC Campus, E-28692 Villanueva de la Ca\~nada, Madrid, Spain}
\author{Mariela Mart\'inez-Paredes}
\affil{Korea Astronomy and Space Science Institute 776, Daedeokdae-ro, Yuseong-gu, Daejeon, Republic of Korea (34055)}
\author{Donaji Esparza-Arredondo}
\affil{Instituto de Radioastronom\'ia y Astrof\'isica (IRyA-UNAM), 3-72 (Xangari), 8701, Morelia, M\'exico}
\author{Natalia Osorio-Clavijo}
\affil{Instituto de Radioastronom\'ia y Astrof\'isica (IRyA-UNAM), 3-72 (Xangari), 8701, Morelia, M\'exico}

\begin{abstract}

Thanks to ground-based infrared and sub-mm observations the study of the dusty torus of nearby AGN has greatly advanced in the last years. With the aim of further investigating the nuclear mid-infrared emission of the archetypal Seyfert 2 galaxy NGC 1068, here we present a fitting to the N- and Q-band Michelle/Gemini spectra. 
We initially test several available SED libraries, including a smooth, clumpy and two phase torus models, and a clumpy disk plus wind model.
We find that the spectra of NGC\,1068 cannot be reproduced with any of these models. Although, the smooth torus models describe the spectra of NGC\,1068 if we allow to vary some model parameters among the two spectral bands. Motivated by this result, we produced new SEDs using the radiative transfer code {\sc SKIRT}. We use two concentric tori that allow us to test a more complex geometry. We test different values for the inner and outer radii, half opening angle, radial and polar exponent of the power-law density profile, opacity, and viewing angle. Furthermore, we also test the dust grains size and different optical and calorimetric properties of silicate grains. The best fitting model consists of two concentric components with outer radii of 1.8 and 28\,pc, respectively. We find that the size and the optical and calorimetric properties of graphite and silicate grains in the dust structure are key to reproduce the spectra of NGC\,1068. A maximum grain size of 1\,$\mu m$ leads to significant improvement of the fit. We conclude that the dust in NGC\,1068 reaches different scales, where the highest contribution to the mid-infrared is given by a central and compact component. A less dense and extended component is present, which can be either part of the same torus (conforming a flared disk) or can represent the emission of a polar dust component, as already suggested from interferometric observations. 

\end{abstract}

\section{Introduction}\label{sec:intro}

According to the unified model of active galactic nuclei \citep[AGN][]{Antonucci93, Urry-Padovani95}, a dust structure surrounding the central engine and causing obscuration is the angular-stone to explain the different types of AGN observed. This structure, commonly called the dusty torus, is located at a few pc from the supermassive black hole (SMBH). However, the physical details of this structure, such as the dust characteristics (both in terms of chemical composition and of grain size distribution) and its geometrical distribution, are still quite poorly understood \citep[see][for a review]{Ramos-Almeida17}.

One way to study the dusty torus in AGN is throughout the fitting of the spectral energy distribution (SED) to dust models using different distributions and/or chemical compositions. Initially, for the sake of simplicity, most authors used smooth dust distributions using radial and vertical density profiles \citep{Pier92,Granato94,Efstathiou95,Van_Bemmel03,Schartmann05}. Other authors have developed radiative transfer models to reproduce geometries where the dust is distributed in clouds \citep{Nenkova08A, Nenkova08B, Hoenig10B}. This is the so-called clumpy distribution. A mix of smooth plus clumpy distributions has also been proposed \citep{Stalevski12,Siebenmorgen15}. More recently, a more complex scenario has been proposed to explain the infrared nuclear emission of Seyfert galaxies. \cite{Hoenig17} produced a model that includes a compact, geometrically thin disk in the equatorial region of the AGN, and an extended, elongated polar structure \citep[see also][]{Stalevski19}. 
The interpretation of high spatial resolution spectra, achieved by means of SED fitting techniques exploiting theoretical emission models, is the key to an unbiased study of the dust properties in AGN, limiting the effect of the host galaxy emission \citep[][]{Ramos-Almeida09, Ramos-Almeida11, Alonso-Herrero11, Gonzalez-Martin19A}.

NGC\,1068 (D=10.58 Mpc; we used the average distance independent of redshift reported in NED\footnote{The NASA/IPAC Extragalactic Database (NED) is operated by the Jet Propulsion Laboratory, California Institute of Technology, under contract with the National Aeronautics and Space Administration.}) is considered to be the prototype Seyfert 2 galaxy \citep[][]{Bland-Hawthorn97} showing broad lines using polarized light \citep[][]{Miller83,Antonucci85}, where the central source is obscured by dust. Although NGC\,1068 is probably one of the best explored AGN at all wavelengths, there is still controversy on the geometry of the obscurer \citep[e.g.][and references therein]{Gravity20}. This torus has been studied in a large number of works. 

Works at sub-millimeter wavelengths are particularly relevant because they have observed the nuclear molecular gas and dust with unprecedented spatial resolution. \cite{Garcia-Burillo16} used the Atacama Large Millimiter Array (ALMA) to map the molecular and continuum emission from the circumnuclear disk of NGC\,1068 and resolve its dusty torus with a size of $\rm{\sim}$4\,pc. \citet{Imanishi18} found sizes of 13$\rm{\times}$4\,pc and 12$\rm{\times}$5\,pc for the molecular torus as seen by HCN J=3-2 and $\rm{HCO^+}$ J=3-2, respectively. \citet[][]{Lopez-Rodriguez20} detected the polarization signature of the torus by means of magnetically aligned dust grains emission. They find that the torus is inhomogeneous and turbulent. Through the HCN (J=3-2) transition, \cite{Impellizzeri19} identified two disk counter-rotating, an inner disk spanning $0.5\lesssim r \lesssim1.2$ pc, and an outer disk extending upto $\sim$7\,pc. Indeed, the inner disk seems to be linked to the kinematics of the maser spots, which are located in a rotating disc with inner radius of $\rm{\sim}$0.65\,pc and outer radius of $\rm{\sim}$1.1 pc \citep[][]{Greenhill97}, which traces the outer, colder part of the accretion disk. \cite{Garcia-Burillo19} also found that the molecular torus has a radial stratification extending over a range of 10-30 pc, since different tracers shows different sizes: the $\rm{HCO^+(4-3)}$, CO(2-1) and CO(3-2) with full size of $11$, $26$ and $28$\,pc, respectively.

At mid-infrared (MIR) wavelengths several studies have also tried to infer the properties of the dust in NGC\,1068. Early works already showed a complex and extended morphology at MIR \citep{Cameron93,Bock00}, which were latter on confirmed by interferometry \citep{Wittkowski04,Jaffe04,Lopez-Gonzaga14} and  spectroscopy \citep{Mason06, Raban09}. SED fitting including MIR emission has also been very useful. For instance, \citet{Lopez-Rodriguez18} used SOFIA, infrared and sub-millimeter observations in order to characterize the emission and distribution of the dust in NGC\,1068, which was firstly studied by \citet{Alonso-Herrero11}. They found that a clumpy torus is able to reproduce the observed emission, although with some discrepancies mainly at short wavelengths. More recently \cite{Pasetto19} fitted the SED of NGC\,1068 to MIR spectra (N- and Q-band) using a smooth torus model finding that the torus has a more complex structure, since they are not able to fit both spectral bands with the same values for the parameters of the torus model. Note that in these works only one or two torus models were used for the SED fitting. 

The aim of this work is to reproduce the 7-23\,$\rm{\mu m}$ MIR spectra of NGC\,1068 (divided into N- and Q-band spectra at 7-13 and 18-23$\rm{\mu m}$, respectively). This kind of work can only be done for NGC\,1068 because is the only AGN with both N- and Q-band ground based observations and, therefore, with enough spatial resolution to isolate the nuclear dust continuum emission from other dust contributors. Future James Web Space Telescope (\emph{JWST}) observations will allow this kind of studies for dozen of nearby AGN. Thus, this work also aims to refine the SED fitting technique in preparation for oncoming \emph{JWST} observations. For that purpose we explore here available models and create new SEDs. In particular we test four different torus models to reproduce the SED of NGC\,1068: the smooth torus model by \cite{Fritz06}, the clumpy torus model by \cite{Nenkova08B}, the two phase torus model by \cite{Stalevski16} and the clumpy disk plus wind model by \cite{Hoenig17}. We confirm the complex dust distribution needed to reproduce the MIR spectra by producing new synthetic SEDs using the 3D Monte Carlo radiative transfer code {\sc SKIRT} \footnote{https://www.skirt.ugent.be}. 

The paper is organized as follows. Section \ref{sec:Torus_models} gives a brief summary of the available dusty models used along this paper. Section\,\ref{sec:Spectral_fitting} describes the spectral fitting to the data to existing models. We explore the dust properties using the radiative transfer code {\sc SKIRT} in Section\,\ref{sec:SKIRT} and the discussion about these results are in Section\,\ref{sec:Discussion}. Finally, in Section\,\ref{sec:Summary} we summarize our main results. 

\begin{figure}[!t]
\begin{center}
\includegraphics[width=1.0\columnwidth]{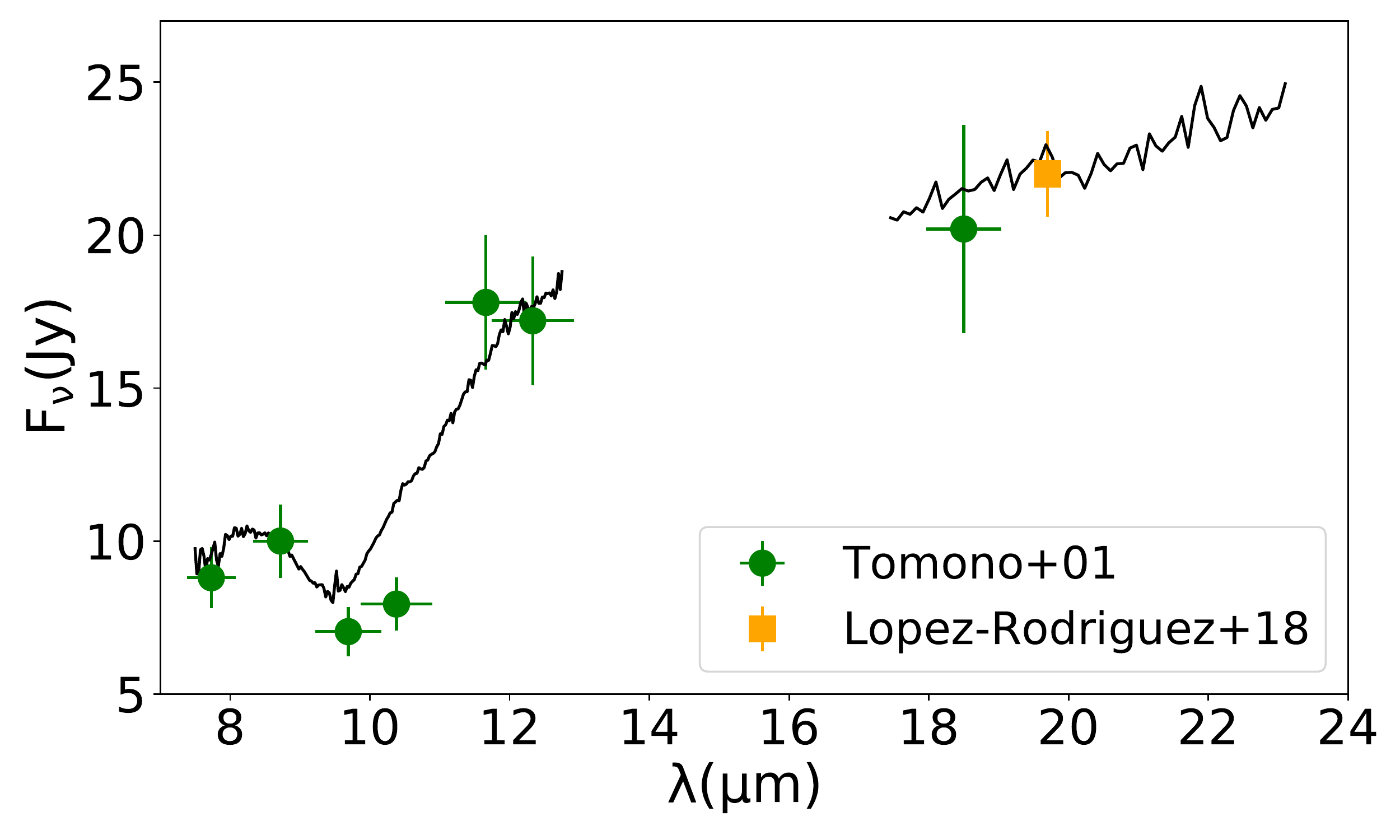}
\caption{N- and Q-band spectra of NGC\,1068 (black lines). The green circles and orange square symbols are the photometry of \cite{Tomono01} and \cite{Lopez-Rodriguez18} respectively.}
\label{fig:data_NGC1068}
\end{center}
\end{figure}

\begin{table*}
\scriptsize 
\begin{center}
\begin{tabular}{ l l c c c}
\hline \hline
Model & Parameter & Dust & Dust & Grain \\ 
& & distribution & composition & size ($\rm{\mu m}$) \\
\hline
\citet{Fritz06} & Viewing angle toward the torus, i & Smooth & Silicate & Silicate:\\
& Half-opening angle of the torus, $\sigma$ & torus & $\rm{\&}$ & 0.025-0.25\\
& Index of the logarithmic azimuthal density distribution, $\Gamma$ & & graphite & Graphite: \\
& Index of the logarithmic radial density distribution, $\beta$ & & & 0.005-0.25 \\
& Ratio between the external and internal radius, Y & \\
& Edge-on optical depth at $\rm{9.7\mu m}$, $\rm{\tau_{9.7\mu m}}$ & \\
\hline
\citet{Nenkova08B} & Viewing angle toward the torus, i & Clumpy & Standard ISM & Silicate:\\
& Number of clouds in the equatorial plane, N & torus & & 0.025-0.25 \\
& Half-opening angle of the torus, $\sigma$ &  &  & Graphite: \\
& Ratio between the external and internal radius, Y & & & 0.005-0.25\\
& Slope of the radial density distribution of clouds, q & & & \\
& Optical depth of the individual clouds, $\rm{\tau_{\nu}}$ & \\
\hline
\citet{Stalevski16} & Viewing angle toward the torus, i & Smooth & Silicate & Silicate:\\
& Half-opening angle of the torus, $\sigma$ & $\rm{\&}$ & $\rm{\&}$ & 0.025-0.25\\
& Index of the logarithmic azimuthal density distribution, p & clumpy & graphite & Graphite: \\
& Index of the logarithmic radial density distribution, q & torus & & 0.005-0.25 \\
& Ratio between the external and internal radius, Y & \\
& Edge-on optical depth at $\rm{9.7\mu m}$, $\rm{\tau_{9.7\mu m}}$ & \\
\hline
\citet{Hoenig17} & Viewing angle toward the torus, i & Clumpy & Standard ISM & Standard:\\
& Number of clouds in the equatorial plane, $\rm{N_0}$ & disk & ISM large & 0.025-0.25\\
& Index of the radial distribution of clouds, a & $\rm{\&}$ & & Large:\\
& Half-opening angle of the wind, $\rm{\theta}$ & outflow & & 0.1-1\\
& Angular width of the walls of the cone, $\rm{\sigma}$ &  \\
& Power law index for dust cloud distribution along the wind, $\rm{a_w}$ &  \\
& Wind-to-disk ratio, $\rm{f_{wd}}$ &  \\
& Optical depth of individual clouds, $\rm{\tau_{cl}}$ (fixed) &  \\
\hline \hline
\end{tabular}
\caption{Summary of used dusty models  described in Sec.\,\ref{sec:Torus_models}. We show the parameters in Col.\,2, dusty distribution in Col.\,3, dust chemical composition in Col.\,4, and grain size in Col.\,5.}
\label{tab:models}
\end{center}
\end{table*}

\section{Torus models} \label{sec:Torus_models}

Here we give a brief summary of the four AGN dust models tested in this paper. The parameters and a sketch of the dust distribution for each model are included in Tab.\,\ref{tab:models} and in Fig.\,\ref{fig:models}, respectively \citep[see also][and references therein]{Gonzalez-Martin19A}:

$\rm{\bullet}$ \underline{Smooth torus model} by \cite{Fritz06}: They use a toroidal geometry, consisting of a flared disc that can be represented as two concentric spheres, delimiting respectively the inner and the outer torus radius, having the polar cones removed (see top-left panel in Fig.\,\ref{fig:models}). For the composition of dust they consider a typical silicate and graphite grain size with radius 0.025-0.25 and 0.005-0.25 $\rm{\mu m}$, respectively in almost equal percentages (52.9\% silicate and 47.1\% graphite).

$\rm{\bullet}$ \underline{Clumpy torus model} by \cite{Nenkova08B}: They use a formalism that accounts for the concentration of dust in clouds, forming a torus-like structure (see top-right panel in Fig.\,\ref{fig:models}). They assume spherical dust grains and a standard Galactic mix of 53\% silicate and 47\% graphite grains.

\begin{figure*}[ht!]
\begin{center}
\includegraphics[width=1.0\columnwidth]{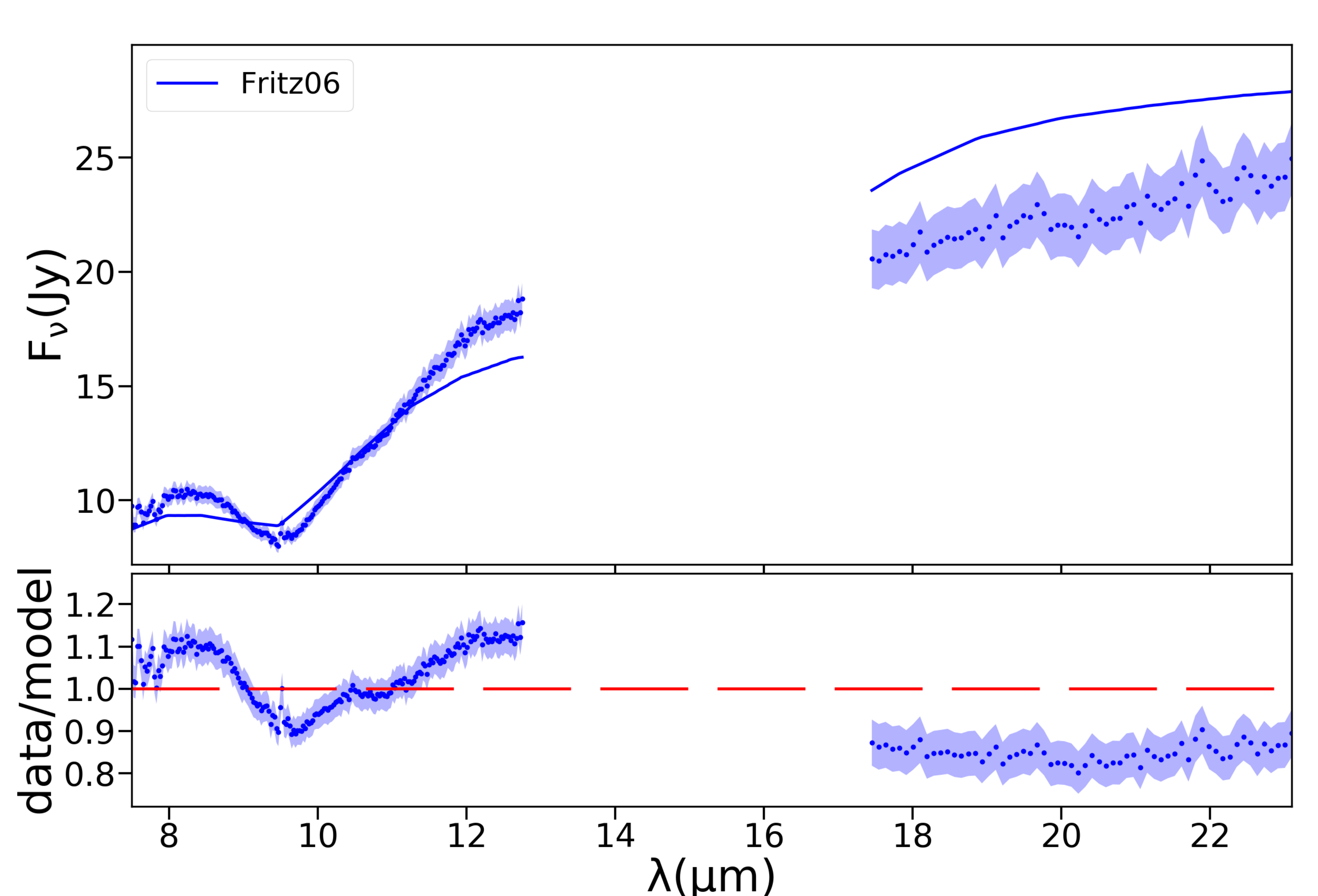}
\includegraphics[width=1.0\columnwidth]{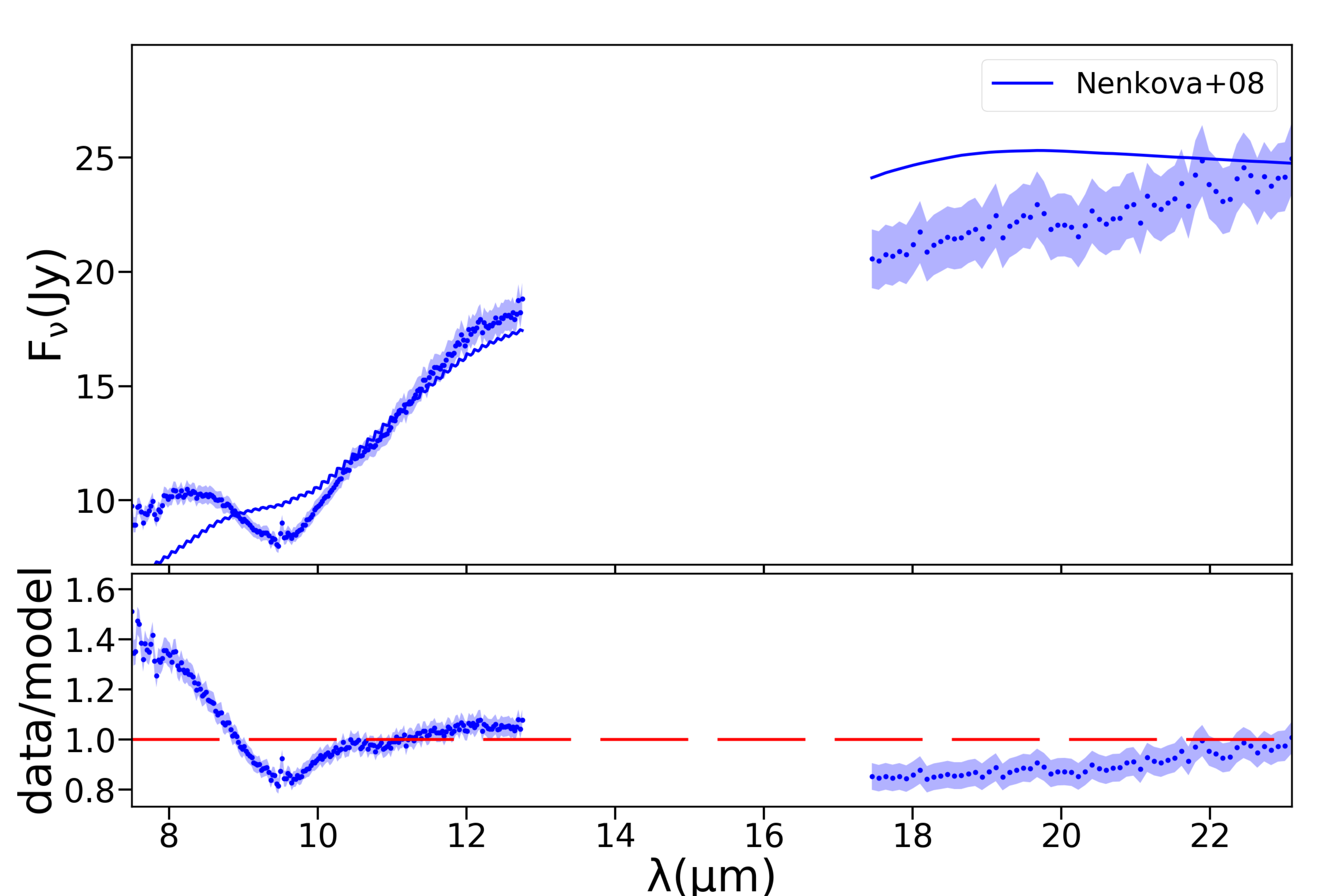}
\includegraphics[width=1.0\columnwidth]{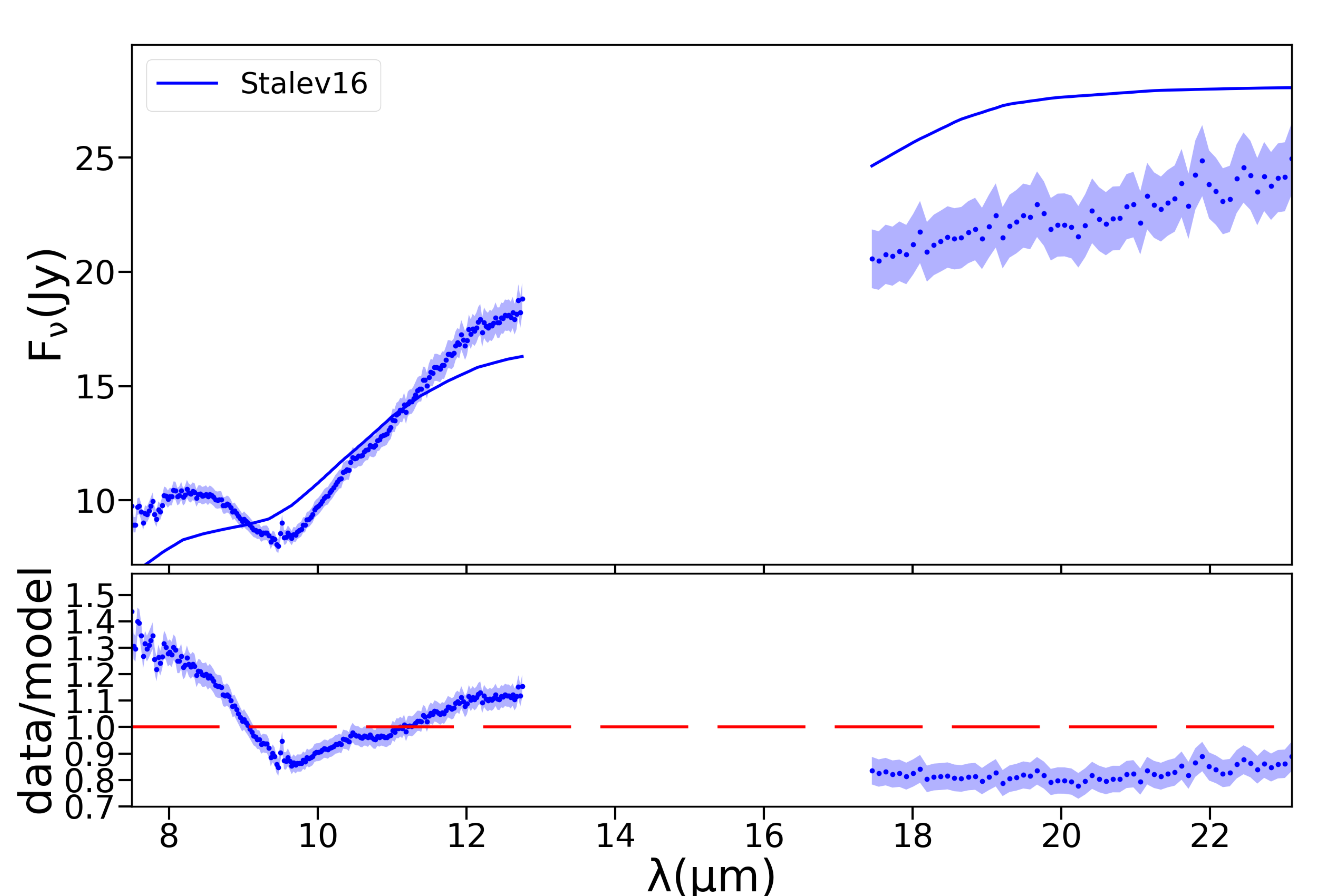}
\includegraphics[width=1.0\columnwidth]{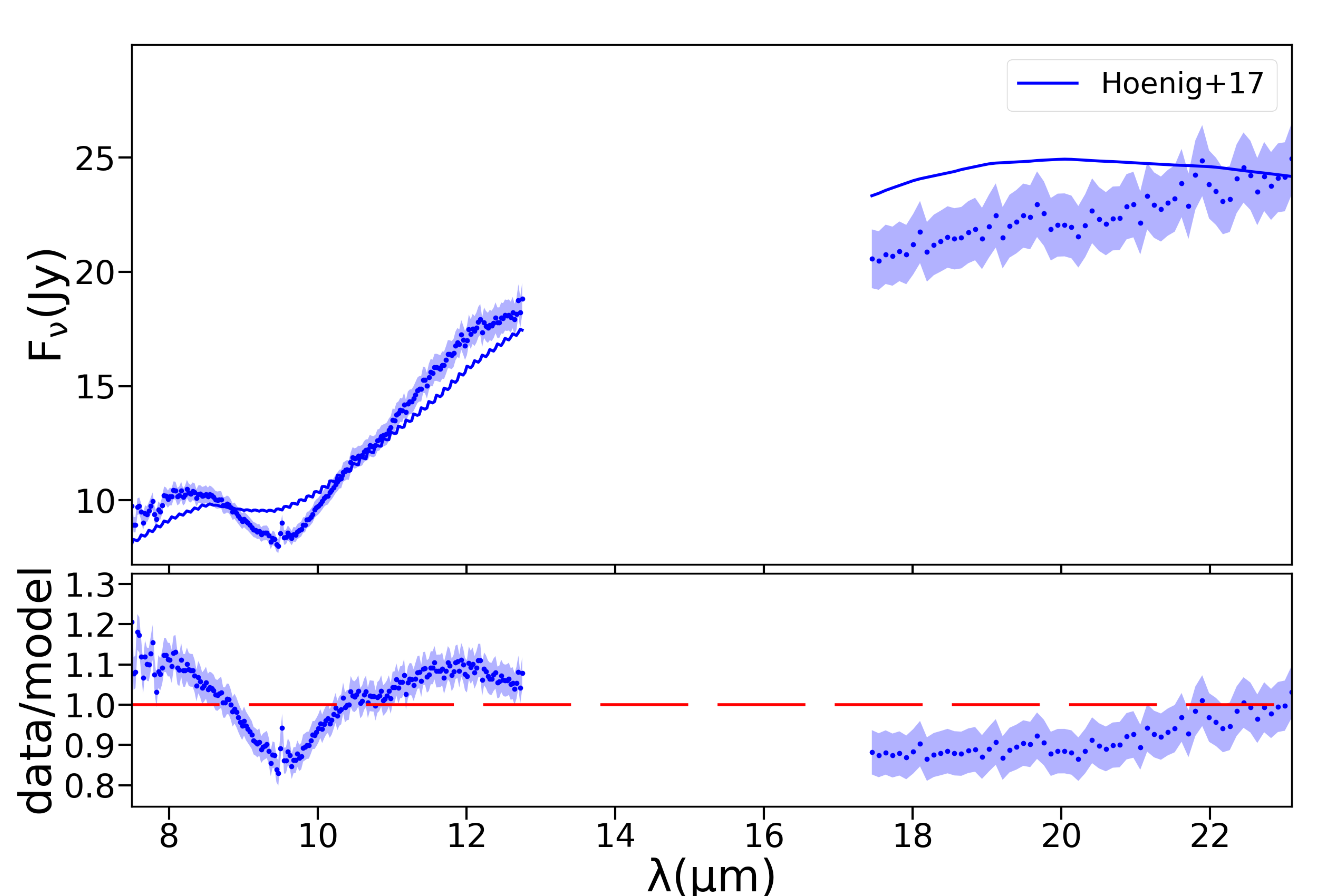}
\caption{MIR spectral fit to NGC\,1068 with the four models fited: the smooth torus model by \citet{Fritz06} (top left), the clumpy torus model by \citet{Nenkova08B} (top right), the two phase torus model by \citet{Stalevski16} (bottom left), and the clumpy disk+wind model by \citet{Hoenig17} (bottom right). We show the best fit (blue solid line) to the data in the top panel and the ratio between model and data in the bottom panel. The dark blue dots show the data and the blue shaded area shows the error on the measurement.}
\label{fig:fits_4models}
\end{center}
\end{figure*}

$\rm{\bullet}$ \underline{Two phase torus model} by \cite{Stalevski16}: They model the dust in a toroidal geometry with a two-phase medium, consisting of high-density clumps embedded in a smooth dusty component of low density (see bottom-left panel in Fig.\,\ref{fig:models}). The dust chemical composition is set to a mixture of silicate and graphite grains. The fraction of total dust mass in clumps compared to the total dust mass is set to 0.97. 

$\rm{\bullet}$ \underline{Clumpy disk and outflow model} by \cite{Hoenig17}: They model the dust in a clumpy disk-like geometry plus a polar hollow cone (see bottom-right panel in Fig.\,\ref{fig:models}). The dust chemical composition is set to only large graphites (0.1-1\,$\rm{\mu m}$) in the outflow. For the disk, the dust composition consists on graphites of 0.1-1\,$\rm{\mu m}$ in the inner part, and a mixture of graphites and silicates of 0.025-0.25\,$\rm{\mu m}$ in the rest of the disk.

\begin{figure*}[!t]
\begin{center}
\includegraphics[width=1.0\columnwidth]{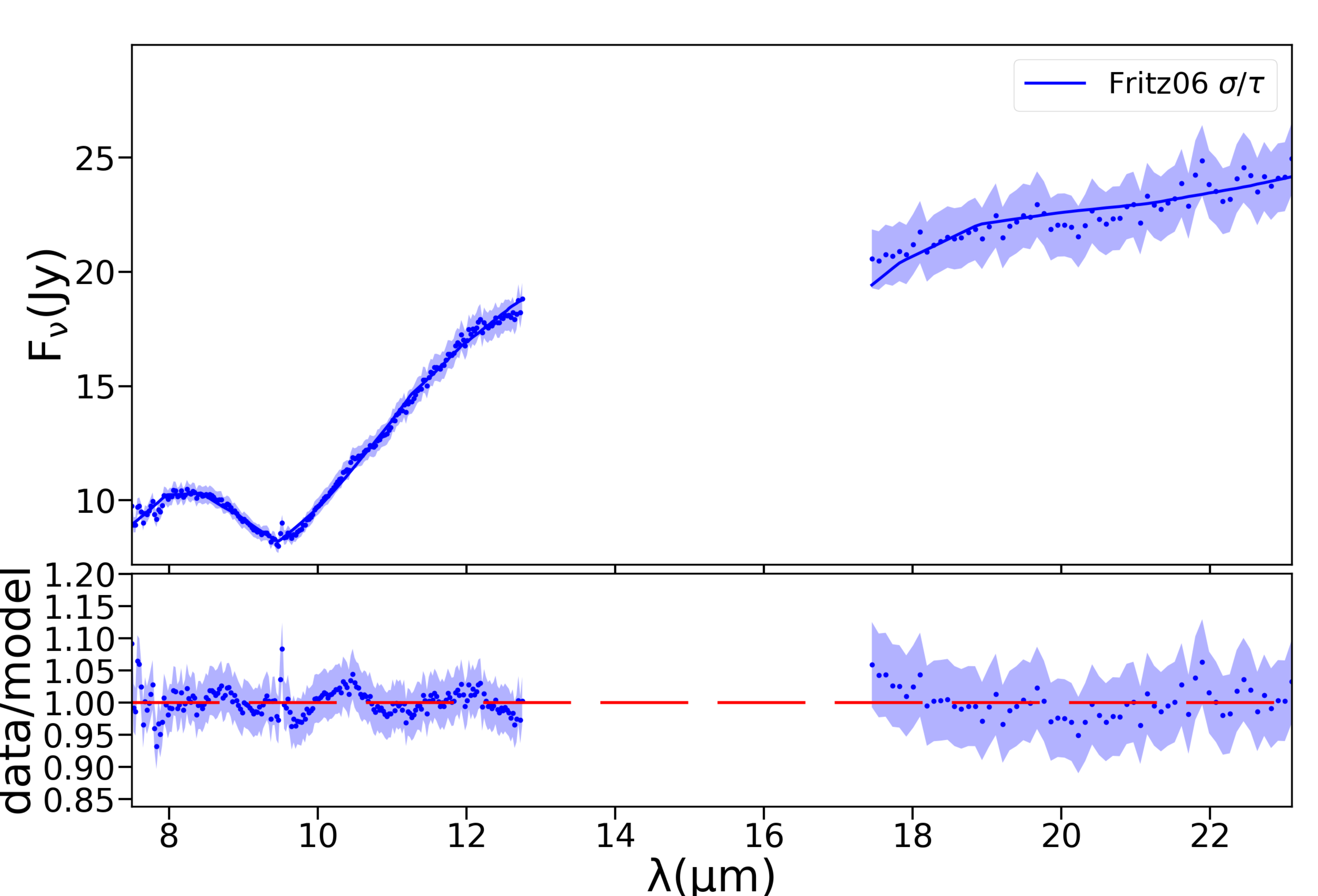}
\includegraphics[width=1.0\columnwidth]{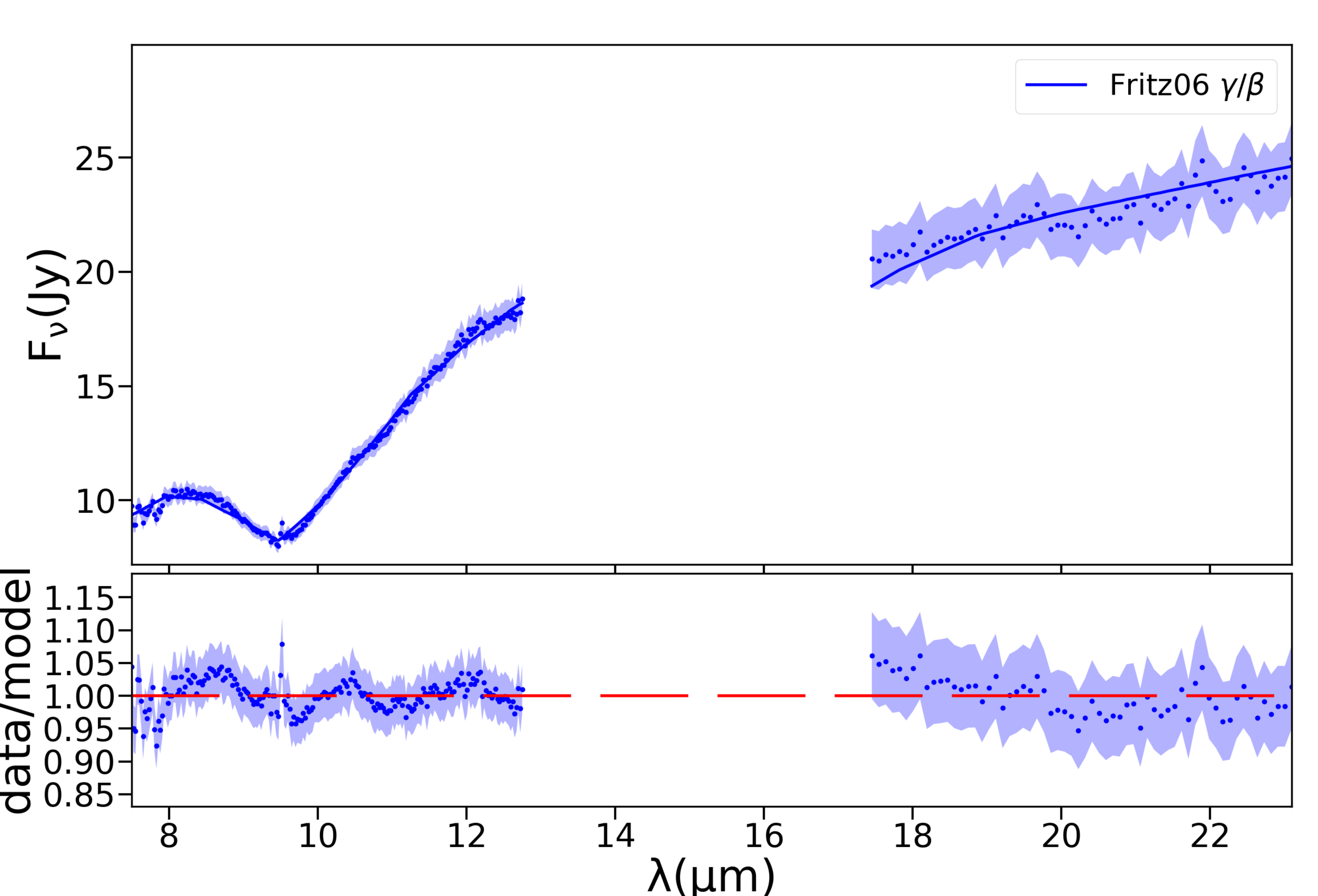}
\includegraphics[width=1.0\columnwidth]{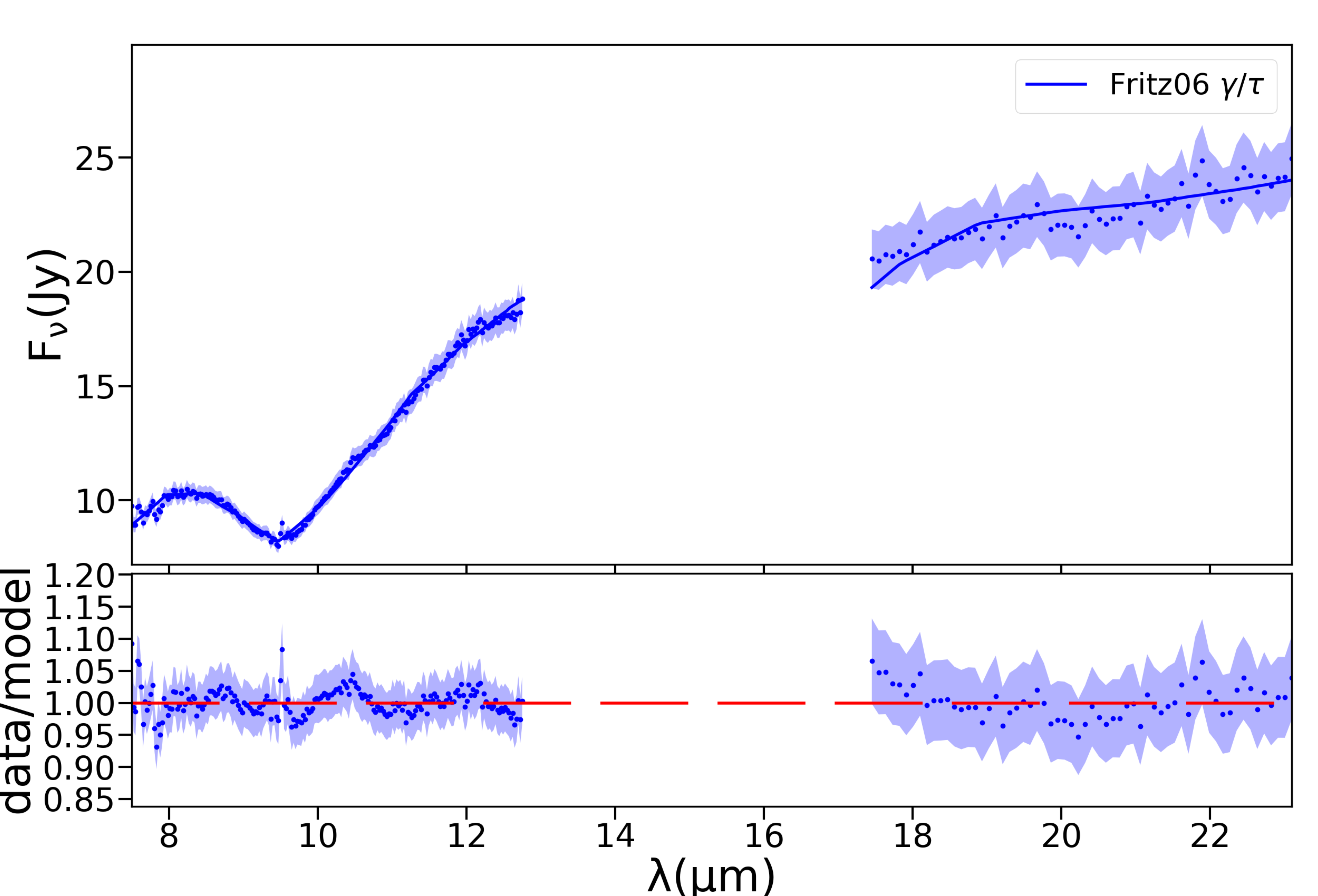}
\includegraphics[width=1.0\columnwidth]{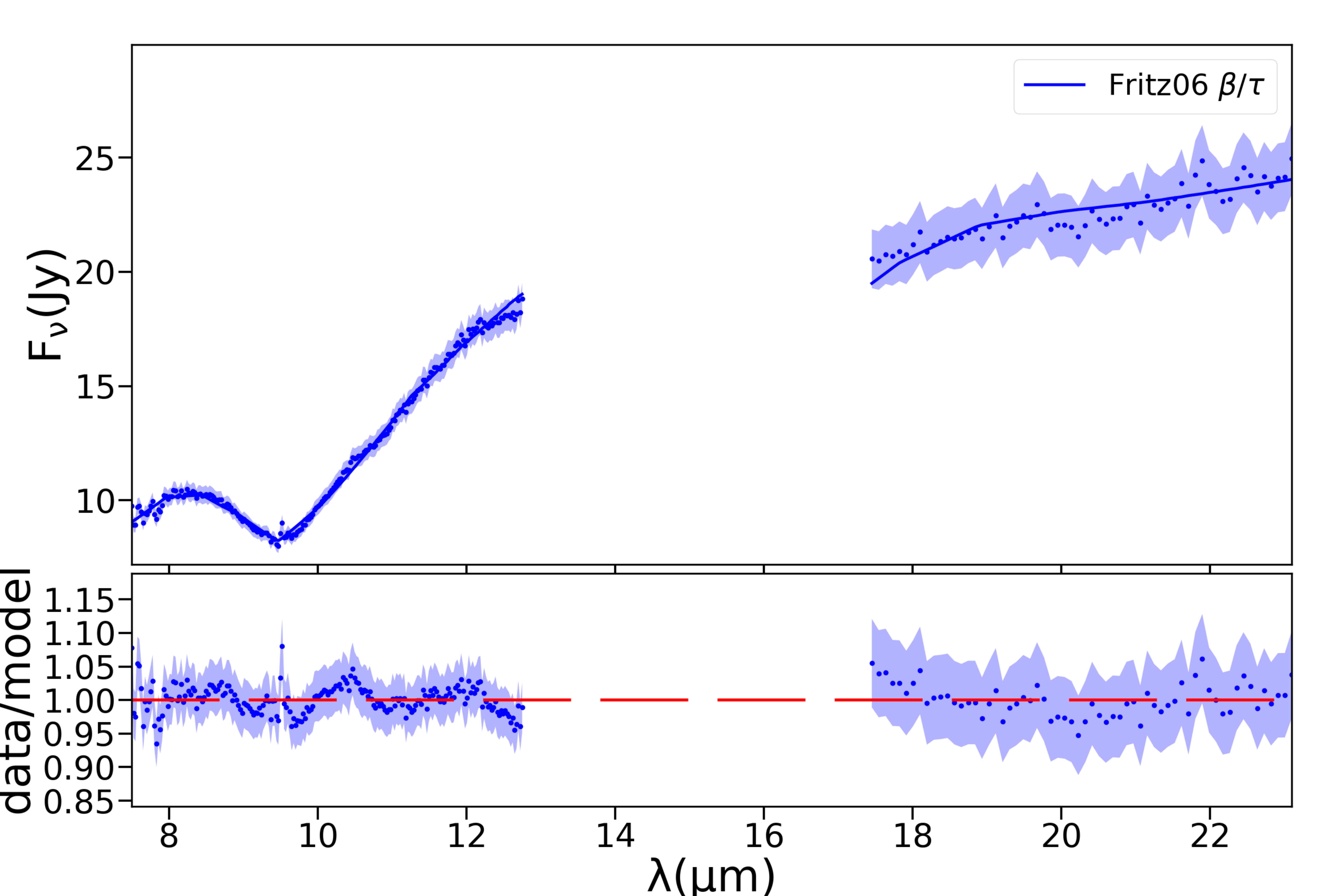}
\caption{Spectral fit to the smooth torus model by \cite{Fritz06} with two parameters unlinked: $\sigma/\tau$ (\textit{top left}), $\gamma/\beta$ (\textit{top right}), $\gamma/\tau$ (\textit{bottom left}) and $\beta/\tau$ (\textit{bottom right}). The description is the same as that given in Fig.\,\ref{fig:fits_4models}.}
\label{fig:Fritz_SEDS}
\end{center}
\end{figure*}

\section{Data and spectral fitting} \label{sec:Spectral_fitting}

We use the N-band (7-13\,$\mu m$) and Q-band (17-23\,$\mu m$) spectra obtained with Michelle spectrometer located in the 8.1m Gemini-North Telescope. This data set was processed for a previous analysis by \cite{Alonso-Herrero11}. They scale the spectra to the corresponding $\rm{0.4^{''}}$ photometric points in order to match the angular resolutions of the imaging and spectroscopic data. The spectra where extracted as point-like sources, following the center and with of the trace of the standard star. PSF and slit-losses corrections were applied to the spectrum. Further details are included in \cite{Alonso-Herrero11}. In Fig.\,\ref{fig:data_NGC1068} we show the N- and Q-band spectra of NGC\,1068. We include the photometric points reported by \cite{Tomono01} using a circular aperture of $\rm{0.4^{''}}$ diameter (similar to the spatial resolution achieved by the ground-based spectra), and a photometric point of \cite{Lopez-Rodriguez18} estimated by using their PSF-scaling method. Note that all but two photometric points (at 9.69 and 10.38 $\rm{\mu m}$) agree with the spectra. A similar figure is shown in \cite{Alonso-Herrero11}. This comparison ensures a proper flux calibration of the N- and Q-band spectra. 

We used the four dust models presented in Section\,\ref{sec:Torus_models} in order to fit the N- and Q-band spectra. Spectral fitting is performed using the {\sc XSPEC} \footnote{http://heasarc.gsfc.nasa.gov/docs/xanadu/xspec/} fitting package, which is a command-driven, interactive, spectral-fitting program within the HEASOFT \footnote{https://heasarc.gsfc.nasa.gov} software. {\sc XSPEC} {\bf{\citep[][]{Arnaud96}}} already includes a large number of incorporated models but new models can be uploaded using the additive table, using the {\sc atable} task. We converted the spectra to {\sc xspec} format in order to upload and fit them to the dust models within {\sc xspec}. In order to assess the goodness-of-fit for each model, we used the reduced $\chi^2$ statistics value. 

We firstly test if both N- and Q-spectra could be fitted with a single SED dust model (Section\,\ref{sec:singleSED}), then we tested more complex SEDs by allowing some parameters to vary among the two bands (Section\,\ref{sec:complex-SED})

\subsection{Single SED dust model}\label{sec:singleSED}

This initial attempt assumes that a single SED is able to simultaneously fit N- and Q-bands. This is the same approximation done in previous works \citep[e.g.][]{Alonso-Herrero11,Garcia-Bernete19}. Figure\,\ref{fig:fits_4models} shows the best fit for the four dust models tested. All the models provide unacceptable fits with reduced $\rm{\chi^2}$, $\rm{\chi_{r}^{2} = \chi^2/dof > 4}$. Among them, the best fit is obtained with the disk+wind clumpy model by \citet{Hoenig17} ($\rm{\chi_{r}^{2} =4.41}$), followed by the smooth torus model by \citet{Fritz06} ($\rm{\chi_{r}^{2}=4.98}$). The other two models provide $\rm{\chi_{r}^{2}>10}$. Tables\,\ref{tab:Fritz_results}-\ref{tab:Hoenig_results} (Col.\,2) in appendix \ref{sec:tables} shows the values obtained for each parameter and the goodness of the fit throughout the $\chi^2$/dof. 

In general all models overestimate the Q-band flux and they struggle to reproduce the 10$\rm{\mu m}$ silicate absorption feature. A clear flux deficit at short wavelengths (below 9$\rm{\mu m}$) is also visible irrespective of the model used.

\subsection{Two SEDs dust model}\label{sec:complex-SED}

Due to the poor fits obtained with the one-SED models, which tend to over-predict the Q-band flux compared with observed spectrum, and show poor agreement in the 10-micron silicate absorption profiles, we propose that a combination of models might provide a better fit, allowing complex dust geometries. Note that this is not an attempt of obtaining a physically motivated fit but to explore the complexity that can achieve a better match to the data. In Section\,\ref{sec:SKIRT} we then use these results to explore new SEDs using the radiative transfer code SKIRT which produces physically motivated SEDs. Therefore, this exploration of complex models is needed to obtain the initial guess for the parameters  in the radiative transfer modelling. This was already explored by \citet{Pasetto19} by unlinking the slope of the azimuthal distribution of dust, $\rm{\gamma}$, and the edge-on optical depth at $\rm{9.7\mu m}$, $\rm{\tau_{9.7\mu m}}$, for the smooth model presented by \citet{Fritz06}. This provided a much better fit in \citet{Pasetto19}. Here we perform a systematic analysis using the four AGN dust models described in Section\,\ref{sec:Torus_models}.

Using the best fit from each model (Sec.\,\ref{sec:singleSED}), we fit separately the N- and Q-band spectra, allowing just one parameter to vary between the two fits. In other words, we kept all the parameters fixed during the fit, except for one of them. We found that changing the parameters does not always result in a improvement of the fit. Tables\,\ref{tab:Fritz_results}-\ref{tab:Hoenig_results} (Col.\,3 and onward) in appendix \ref{sec:tables} show the values obtained and the best statistics (i.e. $\chi^2$/dof) when each parameter of the model is untied between bands. 
In order to study if the fit improved by allowing to vary one of the parameters, we used the f-statistic test (f-test). For the smooth torus model by \citet{Fritz06}, the fit improves by untying any parameter. For the clumpy torus by \citet{Nenkova08B}, the two phase torus by \citet{Stalevski16}, and the clumpy disk+wind by \citet{Hoenig17}, we obtain improved spectral fits by untying any parameter except for $Y$, $p$, and $i$, respectively. However, the improvement is not enough to obtain $\rm{\chi_{r}^{2} \leq 2}$ for the clumpy torus by \citet{Nenkova08B} and clumpy disk+wind by \citet{Hoenig17}. For two phase torus by \citet{Stalevski16}, the fit improves with $\rm{\chi_{r}^{2} \leq 2}$ only by untying the half opening angle of the torus $\rm{\sigma}$; and for the smooth torus model by \citet{Fritz06}, the fit improves significantly by untying any parameter except for the viewing angle, $i$ ($\rm{\chi_{r}^{2}=2.17}$). Thus, unlinking almost any parameter improves the final fit, pointing to the complex nature of the source. 


\begin{table*}[ht!]
\scriptsize 
\begin{center}
\begin{tabular}{cccccccccccc}\hline
\multicolumn{12}{c}{Smooth torus model by \cite{Fritz06}}\\\hline
\multirow{2}{*}{Par.} & \multirow{2}{*}{band} & \multicolumn{10}{c}{unlinking}\\
& & $\sigma/\gamma$ & $\sigma/\beta$ & $\sigma/Y$ & $\sigma/\tau$ & $\gamma/\beta$ & $\gamma/Y$ & $\gamma/\tau$ & $\beta/Y$ & $\beta/\tau$ & $Y/\tau$ \\
\hline\hline
\multirow{2}{*}{i} & N & $40\pm0.5$ & $>0.01$ & $>0.01$ & $>0.01$ & $>0.01$ & $>0.01$ & $>0.01$ & $>0.01$ & $40.35\pm^{0.18}_{0.91}$ &$40.0\pm^{0.27}_{2.2}$\\
& Q & - & - & - & - & - & - & - & - & - & - \\
\multirow{2}{*}{$\sigma$} & N & $25.49\pm^{0.69}_{0.51}$ & $>20$ & $33.89\pm^{2.17}_{0.87}$ & $>20$ & $>20$ & $>20$ & $>20$ & $>20$ & $30.42\pm^{2.41}_{1.2}$ &$36.19\pm^{1.93}_{1.83}$\\
& Q & $>20$ & $39.98\pm^{0.6}_{0.5}$ & ${40.08\pm^{1.64}_{0.54}}$ & $>20$ & - & - & - & - & - & - \\
\multirow{2}{*}{$\gamma$} & N & $0.02$ & $0.08\pm0.01$ & $0.17\pm0.6$ & $0.05\pm^{0.03}_{0.02}$ & $2.0\pm^{0.3}_{0.5}$ & $0.05\pm^{0.05}_{0.01}$ & $>0.01$ & $0.18\pm^{0.11}_{0.7}$ & $2.0\pm^{0.01}_{0.08}$ &$1.99\pm^{0.02}_{0.27}$\\
& Q & $<6$ & - & - & - & $>0.01$ & ${5.06\pm^{0.34}_{2.91}}$ & ${3.56\pm^{1.78}_{0.47}}$ & - & - & - \\
\multirow{2}{*}{$\beta$} & N & $>-1$ & $>-1$ & $>-1$ & $>-1$ & $>-1$ & $>-1$ & $>-1$ & $-0.25\pm^{0.02}_{0.11}$ & $>-1$ &$>-1$\\
& Q & - & $>-1$ & - & - & $-0.94\pm0.01$ & - & - & $>-1$ & $-0.82\pm^{0.04}_{0.05}$ & - \\
\multirow{2}{*}{$Y$} & N & $>10$ & $>10$ & $>10$ & $>10$ & $13.26\pm^{0.75}_{0.81}$ & $>10$ & $>10$ & $113.45\pm^{30.39}_{71.94}$ & $134.89\pm^{4.76}_{4.08}$ &$137.0\pm^{5.15}_{4.35}$\\
& Q & - & - & $>10$ & - & - & ${22.31\pm^{3.46}_{3.31}}$ & - & $>10$ & - & $>10$\\
\multirow{2}{*}{$\rm{\tau_{9.7\mu m}}$} & N & $2.0\pm^{0.02}_{0.07}$ & $2.0\pm^{0.02}_{0.05}$ & $1.87\pm^{1}_{0.03}$ & $1.84\pm^{0.12}_{0.1}$ & $2.0\pm^{0.02}_{0.01}$ & $1.84\pm^{0.6}_{0.11}$ & $1.8\pm0.1$ & $1.93\pm0.02$ & $5.98\pm^{0.03}_{0.22}$ &$5.73\pm^{0.28}_{0.18}$\\
& Q & - & - & - & ${0.15\pm^{0.04}_{0.03}}$ & - & - & ${0.28\pm^{0.05}_{0.14}}$ & - & ${0.14\pm^{0.05}_{0.02}}$ & $1.6\pm^{0.9}_{1.0}$\\
\hline
$\chi^{2}/ dof$ & & 94.04/258 & 151.45/258 & 77.76/258 & 67.37/258 & 89.16/258 & 67.75/258 & 68.88/258 & 101.66/258 & 83.82/258 & 68.54/258 \\
\hline
\end{tabular}
\end{center}
\caption{Values of the parameters and statistics obtained with the smooth torus model by \citet{Fritz06} when two parameters are unlinked at the Q-band compared to the N-band. Note that the viewing angle is measured with respect to the equatorial plane.}
\label{tab:Fritz_unlink_two}
\end{table*}


We then test if untying two parameters significantly improves the resulting fit. We discard from the subsequent analysis clumpy torus by \citet{Nenkova08B} and clumpy disk+wind by \citet{Hoenig17} due to the poor spectral fit obtained so far. In general, we consider those parameters that produce a statistical improvement (see above). For the smooth torus model by \citet{Fritz06} we discard the model obtained by untying viewing angle, $i$, because we do not expect two different values for the viewing angle of the torus. In the case of the two phase torus by \citet{Stalevski16}, we only consider the model obtained by untying the half opening angle, $\rm{\sigma}$, because untying the other parameters does not produce a significant improvement. 

We show in Table\,\ref{tab:Fritz_unlink_two} the results for the smooth torus model by \citet{Fritz06} by untying two parameters between N- and Q-bands. We have ten possible combinations of the parameters. We discard six of them because they do not provide physically plausible scenarios (e.g, due to the better resolution of N-band compared to the Q-band, values for $Y$ in N-band larger than the Q-band are not expected). We have four scenarios that significantly improved the fits by untying: 1) $\sigma$ and $\gamma$; 2) $\sigma$ and $\beta$; 3) $\gamma$ and $\tau$; and 4) $\beta$ and $\tau$. Figure\,\ref{fig:Fritz_SEDS} shows the best fit obtained for these four scenarios.  
For the two phase torus model by \citet{Stalevski16}, we discard the combination untying $\rm{\sigma}$ with any other parameter, because any of them results in a physically plausible scenario, since the half opening angle of the torus related to the N-band (whit better resolution) is bigger that the related whit the Q-band (whit worse resolution). Table\,\ref{tab:Stalevski_unlink_sigma} in appendix \ref{sec:tables} shows the results combining the half opening angle, $\sigma$, with the other parameters for the two phase torus model by \citet{Stalevski16}. 

\begin{figure}[ht!]
\begin{center}
\includegraphics[width=1.0\columnwidth]{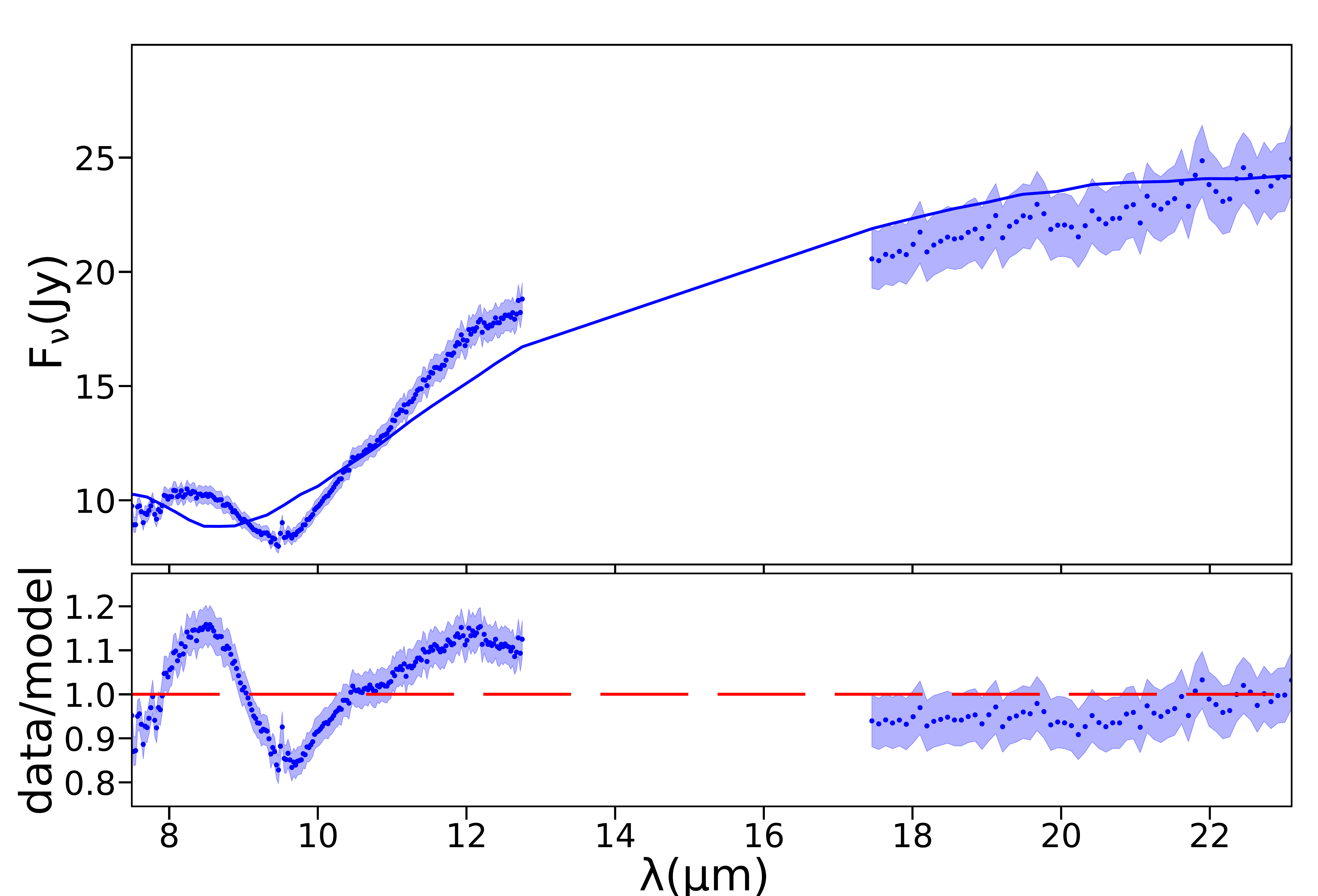}
\includegraphics[width=1.0\columnwidth]{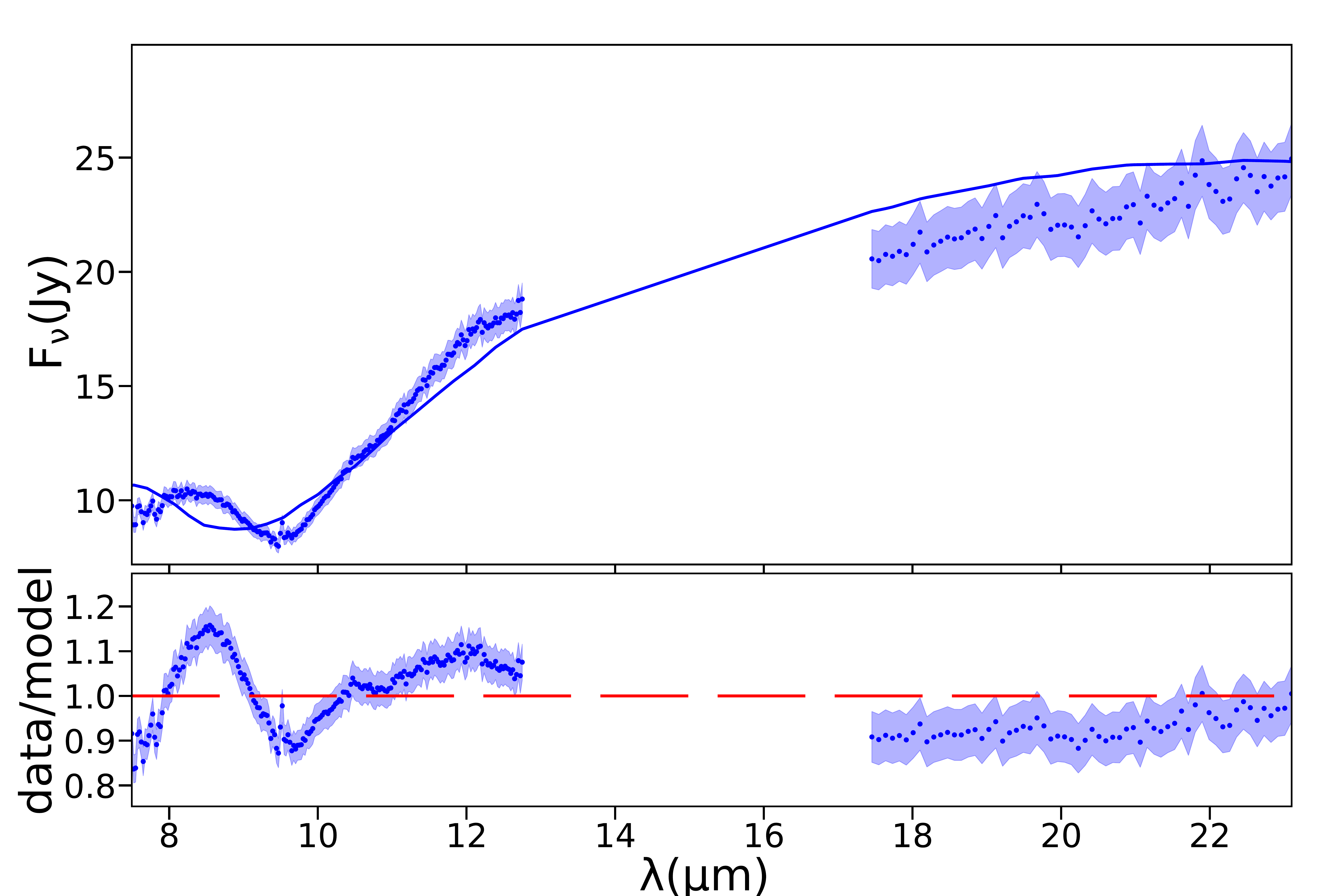}
\caption{(top): Best fit of the N- and Q-band spectra using all the models in Grid A. (bottom): same as Grid A but with a larger range of values for $R_{max}$ (Grid B). In each figure we show the model in blue solid line in the top panel and the ratio between model and data in the bottom panel. The blue shaded area shows the error on the measurement.}
\label{fig:first_second_angle}
\end{center}
\end{figure}

\section{SED simulations with SKIRT} \label{sec:SKIRT}

In order to better explore the complex torus structure of NGC\,1068, we produced synthetic SEDs based on the results from previous section. For this, we used the 3D Monte Carlo radiative SED transfer code {\sc SKIRT} \citep{Baes03, Baes11}. {\sc SKIRT} offers state-of-the-art software for simulating radiative transfer in dusty astrophysical systems. We created grids of the parameters based on the smooth torus geometry used by \cite{Fritz06}, which provides the best fit statistics. We remark that clumpy distributions were not tested to be consistent with the results obtained using the available SED models (see Section\,\ref{sec:Spectral_fitting}). 

We fitted the NGC\,1068 N- and Q-band spectra including the foreground extinction using the extinction law described by \citet{Calzetti00}. We assumed a ratio of total to selective extinction $\rm{R_{V}=}$3.1 and an optical extinction in the range $\rm{A_{V}}$=[0-10] magnitudes. This extinction is applied to each SED to test if additional foreground extinction improves the final fit.

In order to determine the best synthetic SED, we used the $\rm{\chi^2}$ statistics through the standard $\rm{\chi^2_r = \chi^2/dof}$, where dof is the total number of data bins in the spectrum. We also use the bayes factor (see appendix \ref{sec:AIC} for details) to evaluate to what extent a model is better than another one. When the bayes factor is $\rm{\leq 0.01}$, the first model is preferred. The second model is preferred when the bayes factor is $\rm{\geq 100}$.



We initially used a mix of graphite and silicate grains from \citet{Li-Draine01} (already available within {\sc SKIRT}), hereafter identified as $\rm{Graph_L}$ and $\rm{Sil_L}$, respectively. We also selected a smooth distribution of the dust particles covering a minimum grain size of $\rm{size(min)=0.005}$ and a maximum grain size of $\rm{size(max)=0.25}$ microns. Moreover, we assume a percentage within the dust mix of 49\% and 51\% for graphite and silicate grains, respectively, close to the dust mix used by \citet{Fritz06}. Geometry, grain size, composition, and optical depth are further explored in Sections\,\ref{sec:dust geometry}, \ref{sec:dust particle size} and \ref{sec:tau}.

\subsection{Grid A and B: Geometry and optical depth} \label{sec:dust geometry}

Following the results found in Section\,\ref{sec:complex-SED}, we constructed a dust geometry conformed by two tori coexisting in the same plane and we explore the parameters of each of the two tori. We varied the radial power-law exponent, $p$, the polar index, $q$, the half opening angle, $\rm{\sigma}$, the maximum/outer radius, $\rm{R_{max}}$, the edge-on optical depth and $\tau$ between the two tori for this first grid. Note that the symbols of the parameters are according to the SKIRT notation. In order to cover a large range of values and to optimize the computational time (roughly 2-3 days per simulation using up to 30 GB of RAM and 12 cores) 
we impose the following physical restrictions. Firstly, the outer radius should be smaller in one of the tori. Moreover, the radial power-law exponent, $\rm{q}$, the polar index, $\rm{\gamma}$, and the equatorial optical depth, $\rm{\tau_{9.7\mu m}}$, must be different for each torus. Finally, the viewing angle of the system, $i$, is the same for both tori. These restrictions allowed to reduce the number of SED produced by focusing in meaningful scenarios. All together we produce a total of 2376 synthetic SEDs. We call this set of models Grid A. The set of parameters tested are included in Col.\,2 of Table\,\ref{tab:initital_grid}.

The best fit parameters are given in Col.\,3 of Table\,\ref{tab:initital_grid}. 
Note that the suffix $1$ and $2$ is used to discriminate between the two tori. Although the best fit shows a poor statistic ($\rm{\chi^{2}_r=5.27}$), it is as good as a single SED from the smooth torus by \citet{Fritz06} and the clumpy disk-wind model by \citet{Hoenig17}, and already better than the two-phase torus by \citet{Stalevski16} and the clumpy torus model by \citet{Nenkova08B}. 
Fig.\,\ref{fig:first_second_angle} (top panel) shows the best fit for Grid A. The center of the silicate absorption feature appears at shorter wavelengths compared to the data and the slope between N- and Q-band also fails to be reproduced by this SED.


We then create a new grid with the best values obtained previously and a larger range of values for the outer radius of the structure, $R_{max}$, named as Grid\,B in Table\,\ref{tab:initital_grid}. 
We impose again that the outer radius of the structure is at least equal or larger in one of the torus. Notice that we do not explore outer radius greater than 30 pc since they would be spatially resolved with the current data, which is not the case. The total number of SEDs produced was 726, including three viewing angles.
The best fit ($\rm{\chi^2_r=3.74}$, Fig.\,\ref{fig:first_second_angle}, bottom panel) 
is already better than any single SED fit reported in Section\,\ref{sec:Spectral_fitting}. The best fit from the second grid has a bayes factor of $10^{89}$ compared to the previous one. However, the issues found for the first grid remain; i.e, the center of the silicate feature is displaced compared to the spectrum and slope between the N and the Q bands is not well recovered. We also explored a broad range for the viewing angle, although this did not yield better results. Thus, hereinafter we focus our analysis in viewing angles in the range of $i$=[50-75]$\rm{^{\circ}}$, which is also consistent with the type-2 nature of NGC\,1068.

\begin{table*}[ht!]
\scriptsize 
\begin{center}
\begin{tabular}{c|cccc|ccccccccc}\hline  \hline
    & \multicolumn{4}{c|}{Geometry and optical depth (Sect.\,4.1)} & \multicolumn{6}{c}{Grain size and composition (Sect.\,4.2)}  \\ 
 & \multicolumn{2}{c}{Grid A} & \multicolumn{2}{c|}{Grid B} & \multicolumn{2}{c}{Grid C} & \multicolumn{2}{c}{Grid D} & \multicolumn{2}{c}{Grid E} \\
Param. & grid & best-fit & grid & best-fit & grid & best-fit & grid & best-fit & grid & best-fit \\
\hline  \hline
$\rm{i}$ & [60,75,90]$\rm{^{\circ}}$ & 60 & [60,75,90$\rm{^{\circ}}$] & 60 & [50-75$\rm{^{\circ}}$] & 55 & [50-75$\rm{^{\circ}}$] & 60 & [50-75$\rm{^{\circ}}$] & 60 & \\
$\rm{\sigma_1}$ & [20,40,60]$\rm{^{\circ}}$ & $\rm{40^{\circ}}$ & $\rm{40^{\circ}}$ & - & $\rm{40^{\circ}}$ & - & $\rm{40^{\circ}}$ & - & $\rm{40^{\circ}}$ & - \\ 
$\rm{\sigma_2}$ & [20,40,60]$\rm{^{\circ}}$ & $\rm{60^{\circ}}$ & $\rm{60^{\circ}}$ & - & $\rm{60^{\circ}}$ & - & $\rm{60^{\circ}}$ & - & $\rm{60^{\circ}}$ & -\\
$\rm{p_1}$ & [0,1] & 0 & 0 & - & 0 & - & 0 & - & 0 & - \\
$\rm{p_2}$ & [0,1] & 1 & 1 & - & 1 & - & 1 & - & 1 & - \\
$\rm{q_1}$ & [0,3,6] & 3 & 3 & - & 3 & - & 3 & - & 3 & - \\
$\rm{q_2}$ & [0,3,6] & 6 & 6 & - & 6 & - & 6 & - & 6 & - \\
$\rm{R_{max,1}}$ & [2,20] & 2 & [1-4] & 2 & 2 & - & 2 & - & 2 & - \\
$\rm{R_{max,2}}$ & [2,20] & 20 & [1-5,10,20,30] & 30 & 30 & - & 30 & - & 30 & - \\
$\rm{\tau_1}$ & [2,20] & 20 & 20 & - & 20 & - & 20 & - & 20 & - \\
$\rm{\tau_2}$ & [2,20] & 2 & 2 & - & 2 & - & 2 & - & 2 & - \\
$\rm{A_{V}}$ & [0-10] & 0 & [0-10] & 1 & [0-10] & 6 & [0-10] & 3 & [0-10] & 3 & \\
Silicate & $\rm{Sil_{L}}$ & - & $\rm{Sil_{L}}$ & - & $\rm{Sil_{M}}$ & $\rm{Sil_{M}}$ & $\rm{Sil_{M}}$ & - & $\rm{Sil_{M}}$ & - \\
Graphite & $\rm{Graph_{L}}$ & - & $\rm{Graph_{L}}$ & - &  $\rm{Graph_{L}}$ & - & $\rm{Graph_{L}}$ & - & $\rm{Graph_{L}}$ & - \\ 
$\rm{log(size({min}))}$ & -2.3 & - & -2.3 & - & -2.3 & - & -2.3 & - & [-3,-2,-1,0] & -2 \\
$\rm{size({max/min})}$ & 5 & - & 5 & - & 5 & - & [20, 200] & 200 & [10,100,1000] & 100 \\
$\rm{N_{SEDs}}$ & 2376 & & 726 &  & 330 & &  660 & & 2970 \\
$\rm{\chi^2/dof}$ & & 5.27 & & 3.74 & & 4.09 &  & 1.33 & & 1.32  \\
\hline \hline
\end{tabular}
\end{center}
\caption{Parameters tested in the grids described in section\,\ref{sec:dust geometry}. Suffix $1$ and $2$ is used to discriminate between the two tori. In the second grid, for those untested parameters, we use the values obtained with the best previous model. $\rm{Sil_L}$ and $\rm{Graph_L}$: Optical and calorimetric properties of the dust silicate and graphite grains reported by \citet{Li-Draine01}. $\rm{Graph_M}$: Optical and calorimetric properties of the dust graphite grains reported by \citet{Min07} (see text).}
\label{tab:initital_grid}
\end{table*}

\begin{figure}[ht!]
\begin{center}
\includegraphics[width=1.0\columnwidth]{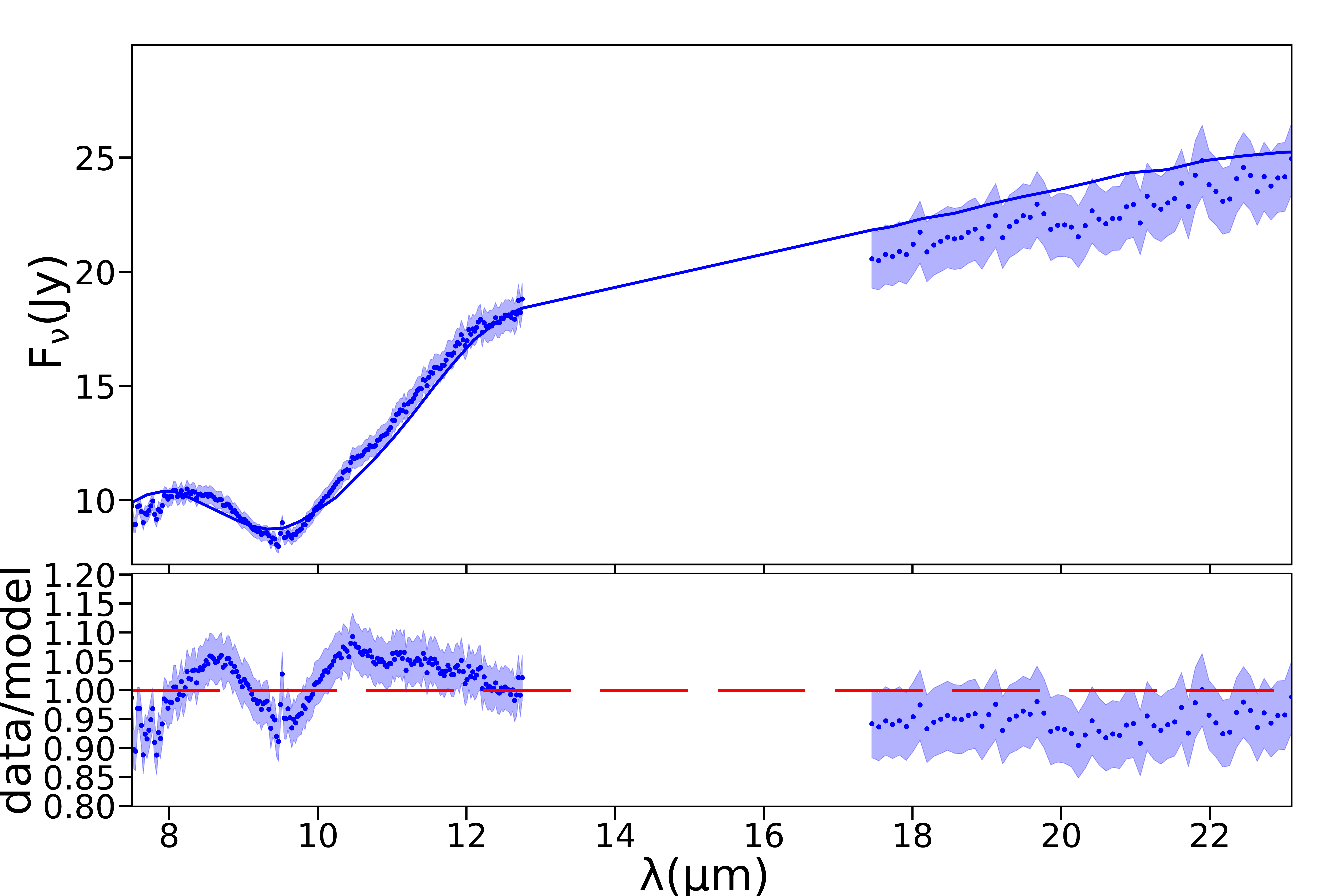}
\caption{Best model for using graphite by \citet{Li-Draine01}, silicate by \citet{Min07} and $\rm{size_1=size_2}$ = 0.005-1\,$\rm{\mu m}$ (Grid E). The description in this figure is the same as that reported in Fig.\,\ref{fig:first_second_angle}. }
\label{fig:grainsizes}
\end{center}
\end{figure}

\subsection{Grids C, D and E: Dust grain size and composition} \label{sec:dust particle size}

As explained above, one of the main problems of our SED simulations is the spectral shape of the silicate absorption feature at $\rm{\sim 9.7\,\mu m}$. This silicate feature is strongly dependent on the dust composition and size. In particular, grain size is expected to be different to that of the ISM. The inner torus radius versus luminosity relation derived by dust reverberation in AGN requires of an emission region more compact than expected. Having a smaller inner radius implies higher temperature than the silicate sublimation temperature, pointing to the existence of large graphite dust grain near AGN \citep{Kishimoto07}. Indeed, large grains can survive close to the accretion disk, where small grains are more easily destroyed. Moreover, the dense environment of AGN promote dust aggregation, possibly making it more efficient for larger than average dust aggregates to form large dust particles.

Although silicate grains with various compositions display a spectral feature in the 10$\rm{\mu m}$ region due to the Si-O stretching mode, there are differences in the spectral appearance (both the peak and shape) depending on the composition of the silicates \citep[see Fig.\,4 by][]{Min07}. Furthermore, the shape and position of the 10$\rm{\mu m}$ silicate feature also has a strong dependence on grain shape. Indeed, the absorption spectrum caused by homogeneous spherical particles is very different from that caused by other particle shapes, being much larger than the differences due to various non-spherical particles shapes \citep{Min03}. In general the spectral extinction features caused by irregularly shaped particles are much broader and shifted towards the red with respect to those caused by homogeneous spherical particles \citep[see Fig.\,3 by][]{Min07}. In practice, both composition and particle shape affects the scattering and absorption efficiencies, rising an impact on the resulting spectrum.


In the previous simulations we used graphite and silicate grains reported by \citet{Li-Draine01} (called here as $\rm{Graph_L}$ and $\rm{Sil_L}$) and we assumed dust grains with sizes in the range of 0.005-0.25 $\rm{\mu m}$, consistent with most of the AGN dust SED libraries reported in the literature \citep[e.g.][]{Fritz06}. In order to explore the effect of optical and calorimetric properties of the dust, we changed the silicate grains from \citet{Li-Draine01} to those reported by \citet{Min07} and available at {\sc SKIRT} (hereinafter $\rm{Sil_M}$). 


\begin{table*}[ht!]
\scriptsize 
\begin{center}
\begin{tabular}{c|cccc|cc}\hline \hline
& \multicolumn{4}{c|}{Optical depth and dust grain size (Sect.\,4.3)} & \multicolumn{2}{c}{Final grid (Sect.\,4.3)}  \\ Par. & \multicolumn{2}{c}{Grid F} & \multicolumn{2}{c|}{Grid G} & \multicolumn{2}{c}{Grid H} \\
& grid & best-fit & grid & best-fit & grid & best-fit \\
\hline  \hline
$\rm{i}$ & [55-90]$\rm{^{\circ}}$ & 63 & [55-90]$\rm{^{\circ}}$ & 67 & [55-90]$\rm{^{\circ}}$ & 71 \\
$\rm{\sigma_1}$ & $\rm{40^{\circ}}$ & - & $\rm{40^{\circ}}$ & - & [38,40,42]$\rm{^{\circ}}$ & 42 \\
$\rm{\sigma_2}$ & $\rm{60^{\circ}}$ & - & $\rm{60^{\circ}}$ & - & [58,60,62]$\rm{^{\circ}}$ & 58 \\
$\rm{p_1}$ & 0 & - & 0 & - & [0,0.1,0.2] & 0.2 \\
$\rm{p_2}$ & 1 & - & 1 & - & [0.8,1,1.2] & 1 \\
$\rm{q_1}$ & 3 & - & 3 & - & [2.8,3,3.2] & 3.2 \\
$\rm{q_2}$ & 6 & - & 6 & - & [5.8,6,6.2] & 5.8 \\
$\rm{R_{max,1}}$ & 2 & - & 2 & - & [1.8,2,2.2] & 1.8 \\
$\rm{R_{max,2}}$ & 30 & - & 30 & - & [28,30,32] & 28 \\
$\rm{\tau_1}$ & [8-22] & 12 & [11,12,13] & 13 & [9-14] & 12 \\
$\rm{\tau_2}$ & [0.4-2.4] & 0.4 & [0.3,0.4,0.5] & 0.3 & [0.1-0.5] & 0.3 \\
$\rm{A_{V}}$ & [0-10] & 4 & [0-10] & 0 & [0-10] & 2 \\
Silicate & $\rm{Sil_{M}}$ & - & $\rm{Sil_{M}}$ & - & $\rm{Sil_{M}}$ & - \\
Graphite & $\rm{Graph_{L}}$ & - & $\rm{Graph_{L}}$ & - & $\rm{Graph_{L}}$ & - \\ 
$\rm{log(size({min}))}$ & -2.3 & - & [-3,-2.3,-2,-1] & -1 & -1 & - \\
$\rm{size({max/min})}$ & 200 & - & [10,100,200,1000] & 10 & 10 & - \\
$\rm{N_{SEDs}}$ & 19008 & & 14256 & & 288684 \\
$\rm{\chi^2/dof}$ & & 1.04 & & 0.48 & & 0.40 \\
\hline  \hline
\end{tabular}
\end{center}
\caption{Parameter grids tested in Section\,\ref{sec:tau}. Suffix $1$ and $2$ is used to discriminate between the two tori. For Grid F we vary the optical depth in steps of $\rm{\Delta(\tau_1)}$ = 2 and $\rm{\Delta(\tau_2)}$ = 0.2.}
\label{tab:final_grid}
\end{table*}

We initially explored only the effect on the new silicate by \citet{Min07} keeping the size of the particles and the values of the parameters obtained with the best fit in Section\,\ref{sec:dust geometry} (Grid C in Table\,\ref{tab:initital_grid}). The resulting fit shows a statistic of $\rm{\chi^{2}_r=4.09}$, which is worse than the best fit obtained with silicates from \cite{Li-Draine01}. However, the effect of the inclusion of these particles together with a change of grain size results in a significant improvement on the results. We therefore explored different dust grain sizes for both tori, fixing the minimum grain size and allowing to very the maximum grain size (Grid D in Table\,\ref{tab:initital_grid}). 
We obtain a best fit with $\rm{\chi^{2}_r = 1.33}$. This fit has a bayes factor of $\rm{1.8\times10^{140}}$ compared to the previous one. 
Note that we also investigate if keeping the canonical grain size of 0.005-0.25$\rm{\mu m}$ for the largest torus and allowing larger particles of 0.005-1.0\,$\rm{\mu m}$ only for the small torus improves the result. However, this SED produces a slightly worse fit than that of $size_1$=$size_2$=0.005-1.0\,$\rm{\mu m}$.

We also explored whether covering size ranges (i.e. from minimum to maximum gran size distributions) of one, two, and three orders of magnitude for the particles size have an impact on the final fit (Grid E in Table\,\ref{tab:initital_grid}). 
The best statistical fit is found using dust grain sizes of $size_1=size_2$=0.001-1\,$\rm{\mu m}$, $size_1=size_2$=0.01-1\,$\rm{\mu m}$, and $size_1=size_2$=0.1-1\,$\rm{\mu m}$ with $\rm{\chi^{2}_r = 1.52}$, $\rm{\chi^{2}_r = 1.32}$, and $\rm{\chi^{2}_r = 2.07}$, respectively. Thus, in all the cases the best fit is found when the maximum particle grain size is $\rm{\sim 1\mu m}$. Moreover, the preferred minimum grain size is in the range of 0.005-0.01$\rm{\mu m}$. These minimum grain sizes has a bayes factor of $\rm{8.8 \times 10^{-12}}$ and $\rm{8.6 \times 10^{-44}}$ compared to the fit obtained with 0.001\,$\rm{\mu m}$ and 0.1\,$\rm{\mu m}$, respectively. Fig.\,\ref{fig:grainsizes} shows the best resulting fit for Grid E\footnote{Note that minimal differences are found between 0.005-1\,$\rm{\mu m}$ and 0.01-1\,$\rm{\mu m}$.}. It is clear how this fit is better describing the silicate absorption feature and the slope between N and Q bands at the same time.

\subsection{Grids F, G and H: Optical depth and dust grain size} \label{sec:tau}

In order to test the effect of the optical depth, which is sensitive to the other physical parameters, we created a new grid with 19008 SEDs (Grid F in Table\,\ref{tab:final_grid}) exploring a range of values according with the latest best model. The test resulted in a best fit ($\rm{\chi_{r}^{2} = 1.04}$). 
This model has a bayes factor of $\rm{8.7\times10^{59}}$ compared to that obtained previously. Notice that we made these grids with grain sizes of 0.005-1\,$\rm{\mu m}$ (one of the three combinations producing the best fit from the previous tests).

\begin{figure}[ht!]
\begin{center}
\includegraphics[width=1.0\columnwidth]{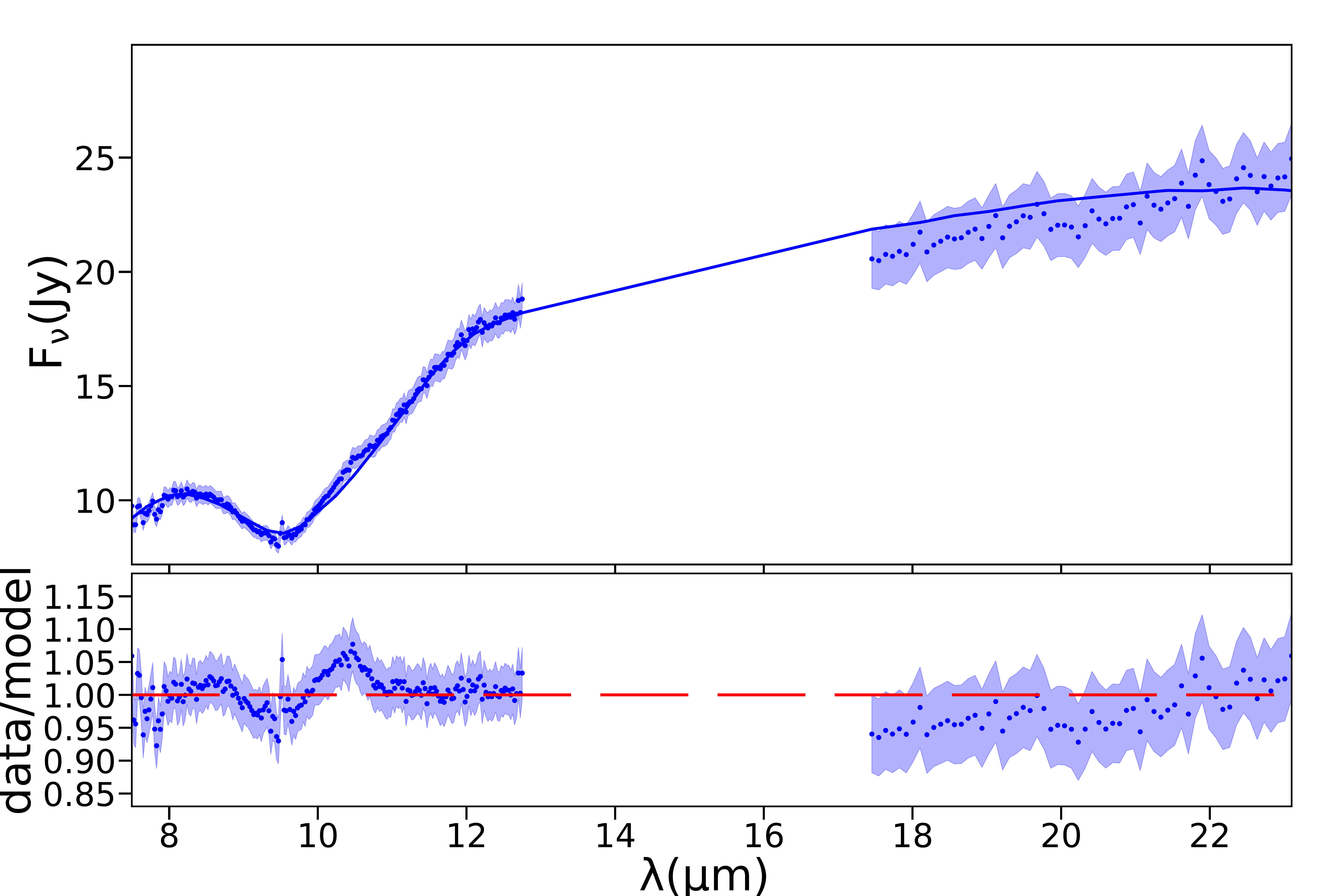}
\caption{Best model exploring the effect on the optical depth, using $\rm{size_1=size_2}$ = 0.1-1\,$\rm{\mu m}$ (Grid G, see Section\,\ref{sec:tau}). The description in this figure is the same as that reported in Fig.\,\ref{fig:first_second_angle}.}.
\label{fig:taus}
\end{center}
\end{figure}

We then create a new grid of SEDs (Grid G in Table\,\ref{tab:final_grid}) exploring the optical extinction $\rm{\tau_{9.7}}$ around the best values from the previous test. 
Moreover, we tested the best grain sizes obtained in the previous section. In total, 14,256 SEDs are crated. We obtained the best fit with $\rm{\chi_{r}^{2} = 0.48}$ using grain sizes of $\rm{0.1-1.0\mu m}$, which is significantly larger than those used in the currently available models. The best fit has a bayes factor of $3.9\rm{2\times10^{32}}$ compared to that obtained in the previous test. Fig.\,\ref{fig:taus} shows the resulting best fit.

Finally, the best fit is expected to depend on the fine tuning of the parameters. For that purpose we explore again a range of parameters around the best fit obtained so far (Grid H in Table\,\ref{tab:final_grid}). 

\begin{figure}[ht!]
\begin{center}
\includegraphics[width=1.0\columnwidth]{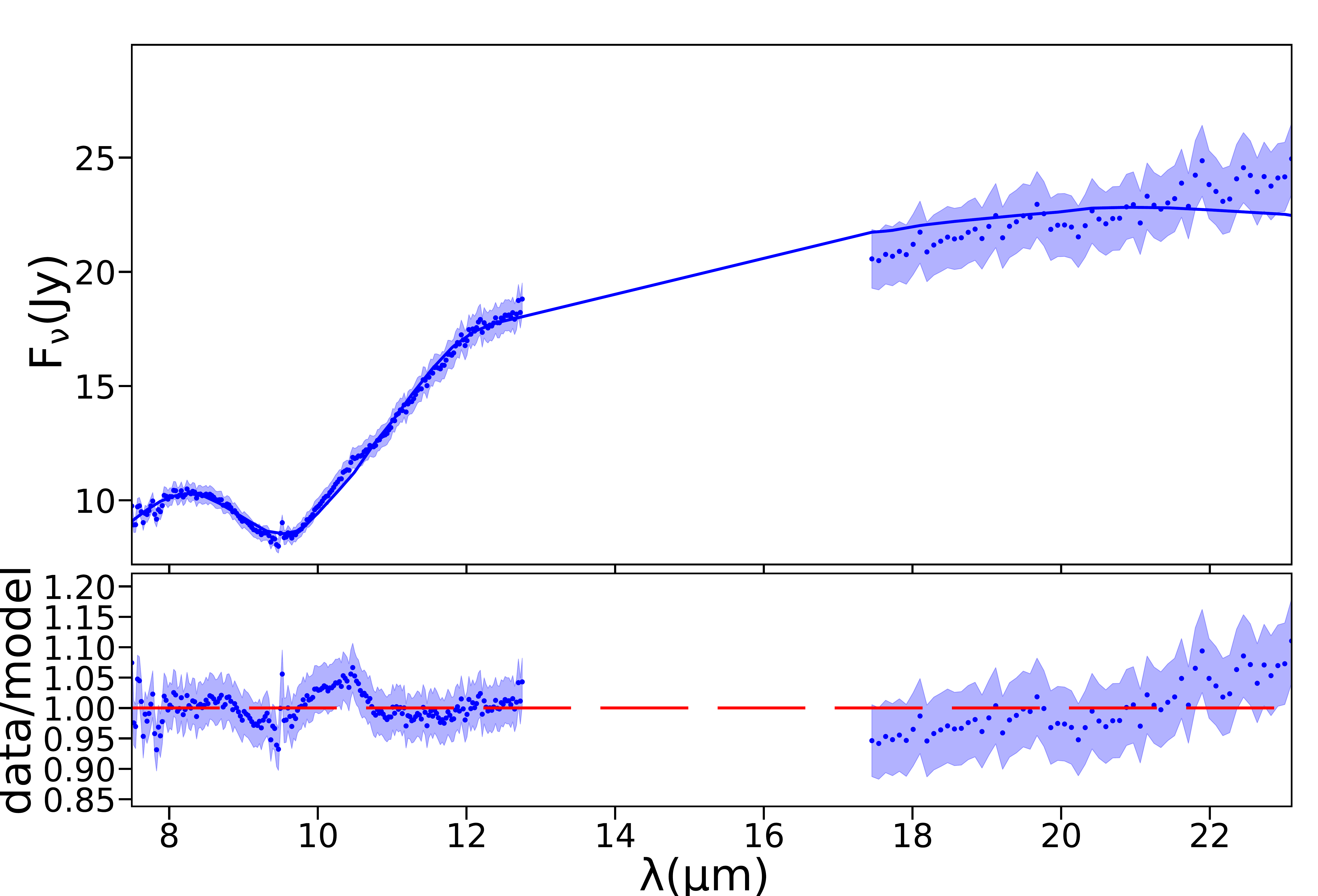}
\caption{Best model obtained after a fine tuning of the parameters (Grid H, see Section\,\ref{sec:tau}). The description in this figure is the same as that reported in Fig.\,\ref{fig:first_second_angle}.}
\label{fig:final}
\end{center}
\end{figure}

The statistic for this fit is $\rm{\chi_{r}^{2} = 0.4}$. This fit has a bayes factor $\rm{4.5\times10^{4}}$ compared to that obtained previously. Fig.\,\ref{fig:final} shows the final best fit obtained.

\begin{figure}[ht!]
\begin{center}
\includegraphics[width=1.0\columnwidth]{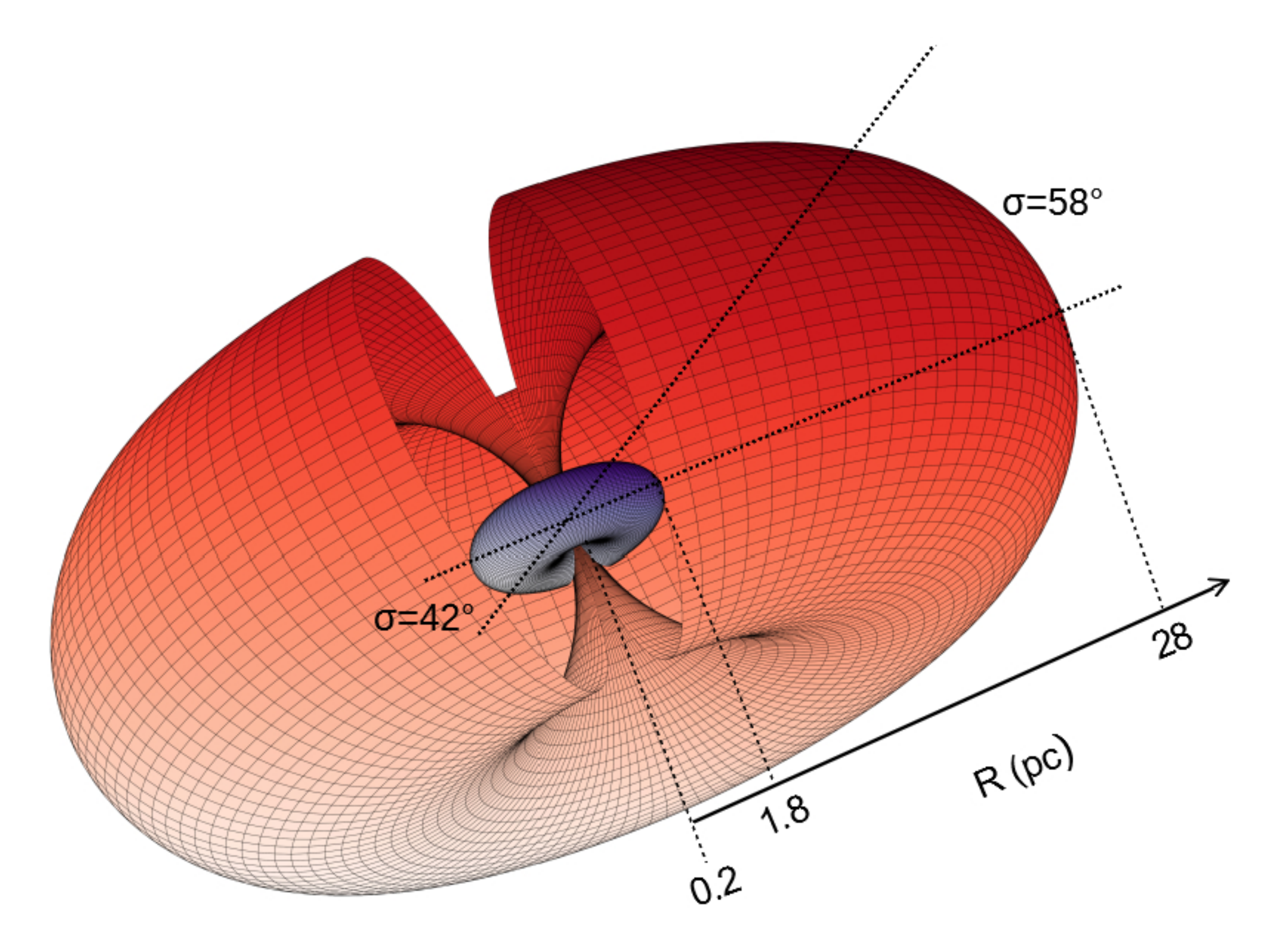}
\caption{Sketch of the dusty distribution of NGC\,1068 obtained through our best model: two concentric tori with inner radius of 0.2\,pc and outer radius of 1.8 and 28\,pc for the smallest (purple) and largest (red) torus respectively. The viewing angle of both tori is $\rm{i=71^{\circ}}$ (shown here perpendicular to the page). We also chose the position angle as $\rm{PA=113^{\circ}}$ and the bottom view of the torus to be consistent with the results presented by \citet{Garcia-Burillo19}. Nothe that we can only distinguish if the torus is  face-on or edge-on with our modelling.}
\label{fig:torus}
\end{center}
\end{figure}

Fig.\,\ref{fig:torus} shows a sketch of the dusty distribution that we obtain through our best model. The smallest (purple) and largest (red) component are two concentric torus with inner radius of 0.2\,pc (according with the dust sublimation radius) and line-of-sight $\rm{i=71^{\circ}}$. The ionization cone in NGC\,1068 as seen in \emph{HST} images is centred around PA = 10$\rm{^\circ}$, while modelling of \emph{HST} spectra based on the kinematics of the gas indicates the ionization cone with an opening angle of 80$\rm{^{\circ}}$ centred around PA = $\rm{30^{\circ}}$ \citep[][]{Das06}, roughly perpendicular to the maser spots. Using SED fitting \cite{Lopez-Rodriguez18} derived a viewing angle of $\rm{i={75^{+8}_{-4}}^{\circ}}$ while \cite{Garcia-Burillo16} found $\rm{i={66^{+9}_{-4}}^{\circ}}$. \cite{Gravity20} found an inclination angle of $\rm{70\pm5^{\circ}}$ through image reconstruction of their interferometric observations of the near-infrared emitting-dust. All of them are consistent with an almost edge-on orientation of the torus, as expected for the type-2 classification of NGC\,1068, which agrees with the value reported in our work. 

Some authors have found that the inclination parameter is very difficult to restrict. \citet{Ramos-Almeida14} concluded that the inclination of the torus is better restricted by using the combination of sub-arsecond resolution near- and mid-infrared. However, at least for NGC\,1068, the inclination angle can be restricted using only mid-infrared spectra. This is due to the availability of spectroscopic data. Indeed, \citet{Gonzalez-Martin19A} found that the viewing angle can be well restricted using only mid-infrared spectroscopy as long as the wavelength range covers at least between 5 to 25$\rm{\mu m}$. Alternatively, the use of combined X-ray and mid-infrared spectra simultaneously has been demonstrated to allow to recover the viewing angle, at least for another type-2 AGN IC\,5063 \citep{Esparza-Arredondo19}.

\section{Discussion} \label{sec:Discussion}

\subsection{Geometry of the dust component}

Early works in the infrared domain as the one presented by \citet{Cameron93} pointed out that complex structures must be present at small scales in NGC\,1068 since the mid-infrared continuum is extended within 1 arcsec. Using higher resolution data, \citet{Bock00} showed there was a central core component ($\rm{<0.2}$\,arcsec), which they associated with the AGN dust torus. However, this core was still not resolved within scales of tens of parsecs. 
\citet{Wittkowski04} and \citet{Jaffe04} presented spatially-resolved images at near- and mid-infrared wavelengths using interferometric techniques. They favor a multi-component model for the dust distribution in NGC\,1068, where part of the flux originates from a hot and small ($\lesssim$ 1 pc) component with also a large warm component (2.1$\rm{\times}$3.4\,pc). Consistent with this result, \citet{Mason06} found that the MIR emission originates in two distinct components. A compact bright source with radius $\rm{<}$ 15 pc, which they identified with the obscuring torus, and a diffuse component with dust in the ionization cones. MIDI observations by \cite{Raban09} found that the N-band mid-infrared emission can be represented by two components. They identified the first component as the inner funnel of the torus, with 0.45$\rm{\times}$1.35\,pc. The second component was identified with the cooler body of the torus, with a size of 3$\rm{\times}$4\,pc. 

From the analysis presented here we conclude that to obtain a good spectral fit using the available AGN dust models it is necessary to consider a two-phase composite geometry for the dusty torus. Indeed, none of the proposed models are able to explain the N- and Q-band spectral range simultaneously (see Section\,\ref{sec:Spectral_fitting}). \citet{Alonso-Herrero11} attempted to simultaneously fit the N and Q spectra in addition to the near-infrared and mid-infrared photometric points. They manage to produce a reasonable fit, using an early version of the clumpy torus model by \citet{Nenkova08B}. The current version of this clumpy torus model did not produce good fits ($\rm{\chi^2_{r} > 4}$). Using the available models, we explored the possibility of adding complexity to them by untying one or two parameters between the fits of the two spectra. We found that the fit improves significantly from the statistical point of view.  
We interpret this as a signal of the complexity of the AGN dust torus in NGC\,1068, consistent with early works.


Through mid-infrared interferometry as well, \cite{Lopez-Gonzaga14} found that the emission of the core of NGC\,1068 can be divided into two distinct regions. However, these two components are not concentric, with one consistent with a hot emission surrounded by warm dust and a large warm diffuse region approximately 7\,pc away from the other. These two components could be associated with two different AGN, conforming a dual or binary AGN. Indeed, \cite{Wang20} suggested that a binary SMBH could support the counter-rotating structures reported by \citet{Imanishi18} and \citet{Impellizzeri19} to explain the molecular gas observations of the torus of NGC\,1068.

If the two AGN are far enough to not significantly influence the heating of dust associated to the other AGN then the total SED should be reproduced with a combination of two AGN dust models. We tested this hypothesis using XSPEC spectral fit software as we did for the single models (i.e. dust + dust models). However, this test did not yield a good spectral fitting ({\bf $\rm{\chi^{2}/dof \geq 5}$}). Instead, allowing parameters of the models to vary independently for the N- and Q-band produced a good fit, as already mentioned above. 

Another possibility is that the two AGN are close enough to heat both AGN dust components. This scenario better match the results by \citet{Wang20} because they assume two AGN with a separation of $\rm{\sim}$0.1\,pc, which is a separation below the expected inner radius of the torus. However, note that this separation is not consistent with the 7\,pc distance between the two dust components found by \citet{Lopez-Gonzaga14}. In the SMBH binary case, the sum of two AGN dust models is not an accurate way to test this scenario and new radiative transfer models are needed including this complexity. We also tested this scenario by producing synthetic SEDs for two non-concentric tori. Details on these new SEDs are reported in Appendix\,\ref{sec:non-concentric tori}. However, the best model obtained is $\sim \rm{1\times10^{12}}$ times worse than the model with two concentric tori reported in our results. Note that the inner radius of the torus is larger than the binary AGN separation given by \citet{Wang20} so this two concentric tori could virtually mimic the dual AGN claimed in their work. Thus, we do not rule out two non-concentric tori associated with a binary SMBH with separations below the inner radius of the torus.   

Studies at other wavelengths can also give information about the complexity of torus. For example, through X-ray data, the reflection component can be used to probe the matter distribution of the gaseous (neutral and distant) torus \citep[see][]{Liu16}. \cite{Bauer15} concluded that a complex reflector structure consisting of multiple components (two nuclear and one extended) is needed to fit the combined \emph{NuSTAR}, \emph{Chandra}, \emph{XMM}-Newton and \emph{Swift}-BAT spectra of NGC\,1068. This is consistent with the two nuclear components shown in our work, although it is worth to notice that the comparison between the X-ray gaseous torus and the mid-infrared dusty torus might be very complex \citep[][]{Esparza-Arredondo21}. 

Through radio-interferometry, molecular line and continuum observations are used for investigating the morphological structure of the torus \citep[][]{Imanishi18}. Although dust continuum from ALMA might have some issues due to jet contamination \citep{Pasetto19,Garcia-Burillo21}. \citet{Garcia-Burillo19} inferred a size of 28\,pc using the CO(3-2) line, which is in agree with the large torus found in our work. This might be explained due to the fact that ALMA could be imaging the cooler (and therefore extended) dust, being the outer layer of the small torus found in our work. In support on the large torus, \cite{Lopez-Rodriguez20} reproduced the 860\,$\mu m$ polarimetric observations of NGC\,1068, using synthetic polarimetric observations generated with the CLUMPY torus model by \cite{Nenkova08B} 
with an outer radius higher than 9\,pc. Furthermore, \cite{Gratadour15} observed the core of NGC\,1068 with the SPHERE instrument on the VLT, using adaptive optics-assisted polarimetric observations in the near-infrared, finding evidence of an extended torus with 15$\rm{\times}$27\,pc. 

The smallest torus with 1.8\,pc (see Fig.\,\ref{fig:torus}) has a high optical depth $\tau_{9.7\mu m}=12$, in the radial direction the dust distribution is almost constant ($\rho \sim r^{-0.2}$), while in the polar direction it decays quickly ($\rho \sim e^{-3.2|cos\Theta|}$). The largest component has a smaller optical depth $\tau=0.3$ compared with that of the smallest torus and the dust density decreases with $r^{-1}$ and $e^{-5.8|cos\Theta|}$ in the radial and polar direction, respectively. Our result might be interpreted as a complex distribution of dust rather than two distinctive components. Our components might be a simplification of a structure where the dust in the outskirts is geometrically thicker than the inner side and the opacity also abruptly falls toward the outskirts.

Alternatively, we could also think that the large torus is just the inner dust from the host galaxy or a polar contribution to the dust, as suggested for other nearby AGN \citep[][]{Hoenig13,Lopez-Gonzaga14,Asmus19} and also for NGC\,1068 \citep[][]{Mason06, Liu19}. This polar dust could produce mostly silicate emission features 
\citep[][]{Hoenig17}. Although not included here, we also produce the SEDs of the two components alone. This polar dust resembles to that of the large torus component modelled in this work due to its low opacity if the inner torus is not presented (although the inner torus gives the highest fraction contribution). Under this latter scenario, the complexity found for NGC\,1068 might be the result of a more dynamic model, as it is the case of the fountain model by \citet{Wada12}. Under this model the radiation from the central source drives the onset of biconical outflows that start forming at the inner region and subsequently propagate outward. Most of this material becomes a truncated wind that back-flows toward the disk plane forming a geometrically thick disk. The small torus found in this work could be related to the thick inner disk which is the origin of the launching wind, while the large torus could be related to the failed wind producing the geometrically thick disk. \citet[][]{Garcia-Burillo19} also found a kinematic model for the molecular torus, where the gas presents circular motions and a fraction of it, inside the torus, is launched as an outflow. In our two component model, the small torus could be related to the torus component in the model by \citet[][]{Garcia-Burillo19}, and the large torus could be related with the outflow component.
As a final remark, in an attempt to find similarities among other astronomical objects, our resulting geometry is not far away from that proposed for protoplanetary disks; a warm zone emitting at around 10$\rm{\mu m}$ and a much colder region emitting longward to 20$\rm{\mu m}$. \citet{Olofsson09} show in their Fig.\,14 an schematic view of the disk in one of these systems, which might imply a grand unification between the dust in proto-planetary disks and that found in AGN.

Note that, if a distance of 14\,Mpc is adopted \citep[commonly used among other studies, e.g.][]{Mason06} instead of the 10.6\,Mpc used in this paper, the luminosity would be higher by a factor of $\rm{\sim}$1.7 and the physical lengths (such as the inner and outer torus radius) would be higher by a factor of $\rm{\sim}$1.3.

\subsection{Dust composition and dust grain size}\label{sec:discussion_grains}

Besides the geometry, an important aspect explored in this work is that SED fitting allows to study the composition of the dust, which cannot be done with other techniques. Unfortunately, available SED libraries has not been produced to explore grain size, size particle distribution or composition so far, although some works have made an effort to explore it \citep[e.g.][]{Hoenig10B}. Thus, new radiative transfer simulations, as those produced in this work, are mandatory to study if dust composition or dust grain size are important to accurately reproduce the shape of the AGN dust continuum. 

In this work, we demonstrate that an important ingredient, to explain the mid-infrared N- and Q-band spectra of NGC\,1068, is the size of the particles. Since, significantly better spectral fits are found if both graphite and silicate grains have grain sizes in the range $\rm{0.1-1\,\mu m}$ using silicates by \citet[][]{Min07}. We also find an important improvement in the best fit model from setting the maximum dust grain size to  1\,$\rm{\mu m}$, while varying the minimum particle size results in marginally change the final SED. This particle size is much larger than that assumed for publicly available AGN dust models since they rely in the results for the ISM \citep[usually 0.005-0.25\,$\mu m$,][]{Fritz06,Nenkova08B,Stalevski16}.

Some authors \citep[eg.][]{Nikuta09} have argued that the 10-micron silicate feature can be explained through a clumpy dust distribution with a standard ISM dust. In section\,\ref{sec:Spectral_fitting}, we tested a clumpy torus model for NGC\,1068, however the silicate absorption profile showed poor agreement. \cite{Feltre12} performed a comparison between the smooth and clumpy models by \cite{Fritz06} and \cite{Nenkova08B}, respectively. They found that the behaviour of the silicate feature at 9.7 $\rm{\mu m}$ is quite distinct between the two models. However, they concluded that such difference arise from the dust chemical composition assumed by the models and not from the smooth or clumpy morphology, in agreement with our findings.


Some works have already explored the effect on the size of the particles in the context of AGN. \citet{Schartmann05} explored models of dusty tori in the AGNs. They tested effects of a broadening of the grain size distribution, spreading the grain size range with grains of 0.005-10 $\mu m$ and 0.001-10 $\mu m$. They found that the differences are nearly negligible for a face-view angle, but for inclination angles, close to edge-on, a reduced relative depth of the silicate feature towards smaller wavelengths is visible. NGC\,1068 has an almost edge-on view of the torus ($\rm{\sim 71^\circ}$) so the importance of the dust grain sizes are justified. Other authors have also found evidence supporting large dust sizes. Through the ratio of the optical extinction in the visual band to the optical depth of the 9.7\,$\rm{\mu m}$ silicate absorption feature, $\rm{A_V/\Delta \tau_{9.7}}$, \citet{Lyu14} obtained a mean ratio of $\rm{A_V/\Delta \tau_{9.7} \approx 5.5}$ from a sample of 110 type 2 AGNs, which is considerably lower than that of the local ISM of the Milky Way ($A_V/\Delta \tau_{9.7} \approx 18.5$), implying that AGN dust grain size could exceed $\sim$0.4\,$\rm{\mu m}$ \citep[][]{Shao17}. 

Although the dependence of the dust sublimation radius on grain size might be complex \citep[see Fig.\,1 in][]{Absil13}, \citet{Kishimoto07} suggested that the dust sublimation radius vary with the square root of the dust size. In the context of debris disks, \citet{Kobayashi11} developed the equation that link the sublimation radius (we expressed it in units of parsecs), the grain size ($\rm{\mu m}$), luminosity (erg/s), and sublimation temperature (K) as follows:
\begin{equation}
    R_{sub} = 0.11 \left(1 + \frac{1}{x}\right)^{1/2} \left(\frac{L_{bol}}{10^{45}}\right)^{1/2} \left(\frac{T_{sub}}{1300}\right)^{-2}
\end{equation}
\noindent where $x\rm{ = 2\pi s_{grain} (T_{sub}/2898K)}$ and $\rm{s_{grain}}$ is the grain size. Therefore, the resulting larger grain sizes, naturally implies dust located closer to the torus. Considering that the AGN dust models use $\rm{s_{grain}=0.25\mu m}$ while we have found $\rm{s_{grain}=1\mu m}$, the largest grains might be located up to a factor of two closer to the accretion disk. However, we tested the inner radius without a significant improvement in the final fit. Conversely, also in the context of  protoplanetary disks, the analysis of the shape and strength of both the amorphous 10$\rm{\mu m}$ feature and the crystalline feature around 23$\rm{\mu m}$ provides evidence for the prevalence of $\rm{\mu m}$-sized (amorphous and crystalline) grains in the upper layers of disks \citep[][]{Olofsson09}. Observations of dust in disks from sub-mm to cm wavelengths have provided strong evidence for grain growth in disks \citep[][]{Testi01,Calvet02}. \citet{Perez12} found that the maximum size of the particle-size distribution increases from sub-mm sizes in the outer disks to mm and cm sizes in the inner disks. Indeed, dust grains in the planet-forming regions around young stars are expected to be heavily processed due to coagulation, fragmentation, and crystallization \citep[][]{Olofsson09}, and similar mechanisms might explain also the large grains found in NGC\,1068. 

\section{Summary}\label{sec:Summary}

We have studied the dusty torus in NGC\,1068 using N- and Q-band Michelle/Gemini spectra. For this purpose we perform the analysis into two steps: (i) we used {\sc XSPEC} spectral fitting package to test already available models and (ii) we used the 3D Monte Carlo radiative spectral energy distribution transfer code {\sc skirt} to build grids of synthetic SEDs, based on the result obtained from the first step. The main results are:

\begin{itemize}
    \item \underline{Available SED models}: Among the available models, the resulting best fit was obtained using the clumpy disk plus wind model by \citet[][]{Hoenig17}. However, the best fit was statistically unsatisfactory. We then explored the possibility of adding complexity to the models by untying one and then two parameters when doing the fits of the N and Q spectra separately. We found that the fit  significantly improves using the smooth torus model by \citet[][]{Fritz06} and four combinations of untied parameters. In this scenario, the emission in the two spectral ranges here considered is dominated by dust with different geometrical locations and distributions. These components are characterized by different values of the equatorial optical depth, the opening angle, and of the parameters regulating the dust density gradients. We interpret these results as a signal of the complexity of the dust in NGC\,1068.
    
    \item \underline{New SEDs with {\sc skirt}}: For the 3D Monte Carlo radiative simulations, we used two concentric tori that allow us to test more complex dusty geometries. The final best fit has the following common parameters for both tori: 1) a fractional contribution for graphite and silicate grains of $\rm{49\%}$ and $\rm{51\%}$, respectively; 2) graphite from \citep[][]{Li-Draine01} and silicate from \citep[][]{Min07}; 3) size of graphite and silicate particles of 0.1-1\,$\rm{\mu m}$; 4) inner radius of both tori of 0.2 pc; 5) viewing angle i=71$\rm{^{\circ}}$; and 6) foreground extinction $A_V=2$ mag. The parameters changing from both tori are: i) the exponent of the power-law describing the radial distribution $\rm{p_{1}}$=0.2 and $\rm{p_{2}}$=1, ii) the exponent of the polar distribution $\rm{q_{1}}$=3.2, and $\rm{q_{2}}$=5.8, iii) the half opening angle $\rm{\sigma_{1}}$=42 and $\rm{\sigma_{2}}$=58, iv) the outer radius $\rm{R_{max,1}}$=1.8 pc and $\rm{R_{max,2}}$=28 pc, and v) the equatorial optical depth to $\rm{\tau_{9.7\mu m, 1}}$=12 and $\rm{\tau_{9.7\mu m, 2}}$=0.3.
    
\end{itemize}

These findings can be interpreted as a compelling evidence for a complex dusty torus for NGC\,1068. We speculate that this can be understood as inner compact disk/torus plus an outer extended torus/wind, conforming either a flared disk or a dynamical fountain model for the dust. Incoming results of the dusty structure of NGC\,1068 using VLTI/MATISSE observations can be compared with the results presented in this work. Furthermore, some mechanism for grain growth need to be claimed to explain the large grains to fit the mid-infrared ground-based spectra of NGC\,1068. Note that, these results were obtained considering a smooth distribution of dust. Therefore, some parameters of the torus likely depend on the assumption of this distribution. In order to consider a similar geometry with other distributions, like clumpy, future efforts should test these scenarios. As a final remark, it has been largely discussed by the community that the complexity of the models, model parameter degeneration, and spatial resolution might be an issue when inferring the properties of the dust using SED fitting techniques \citep[][]{Ramos-Almeida14,Gonzalez-Martin19A}. However, this detailed work demonstrate that useful information can be achieved from SED fitting when the full mid-infrared spectral coverage is available and specific synthetic SEDs are produced to explain the observations.

\section*{Acknowledgements}

We thank to the anonymous referee for his/her useful comments. CV-C acknowledge support from a CONACyT scholarship. We thank the UNAM PAPIIT project IN105720 (PI OG-M). SGB acknowledges support from the research projects PID2019-106027GA-C44 and SGB and AAH from research project PGC2018-094671-B-I00 funded by  MCIN/AEI/ 10.13039/501100011033 and by ERDF A way of making Europe. CRA acknowledges financial support from the European Union's Horizon 2020 research and innovation programme under Marie Sk\l odowska-Curie grant agreement No 860744 (BiD4BESt), from the State Research Agency (AEI-MCINN) and from the Spanish MCIU under grants ``Feeding and feedback in active galaxies" with reference PID2019-106027GB-C42,  ``Quantifying the impact of quasar feedback on galaxy evolution (QSOFEED)", with reference EUR2020-112266, and from the Consejería de Econom\' ia, Conocimiento y Empleo del Gobierno de  Canarias and the European Regional Development Fund (ERDF) under grant with reference ProID2020010105.

\software{XSPEC \citep{Arnaud96}, HEAsoft \citep{Heasarc14}, SKIRT \citep{Baes03, Baes11}, Python/C languaje \citep{Oliphant07}}, NumPy \citep{Harris20}, SciPy \citep{Virtanen20}, and Matplotlib \citep{Hunter07}.


\clearpage
\appendix

\section{Dust models} \label{sec:dust_models}

\begin{figure*}[ht!]
\begin{center}
\includegraphics[width=1.0\columnwidth]{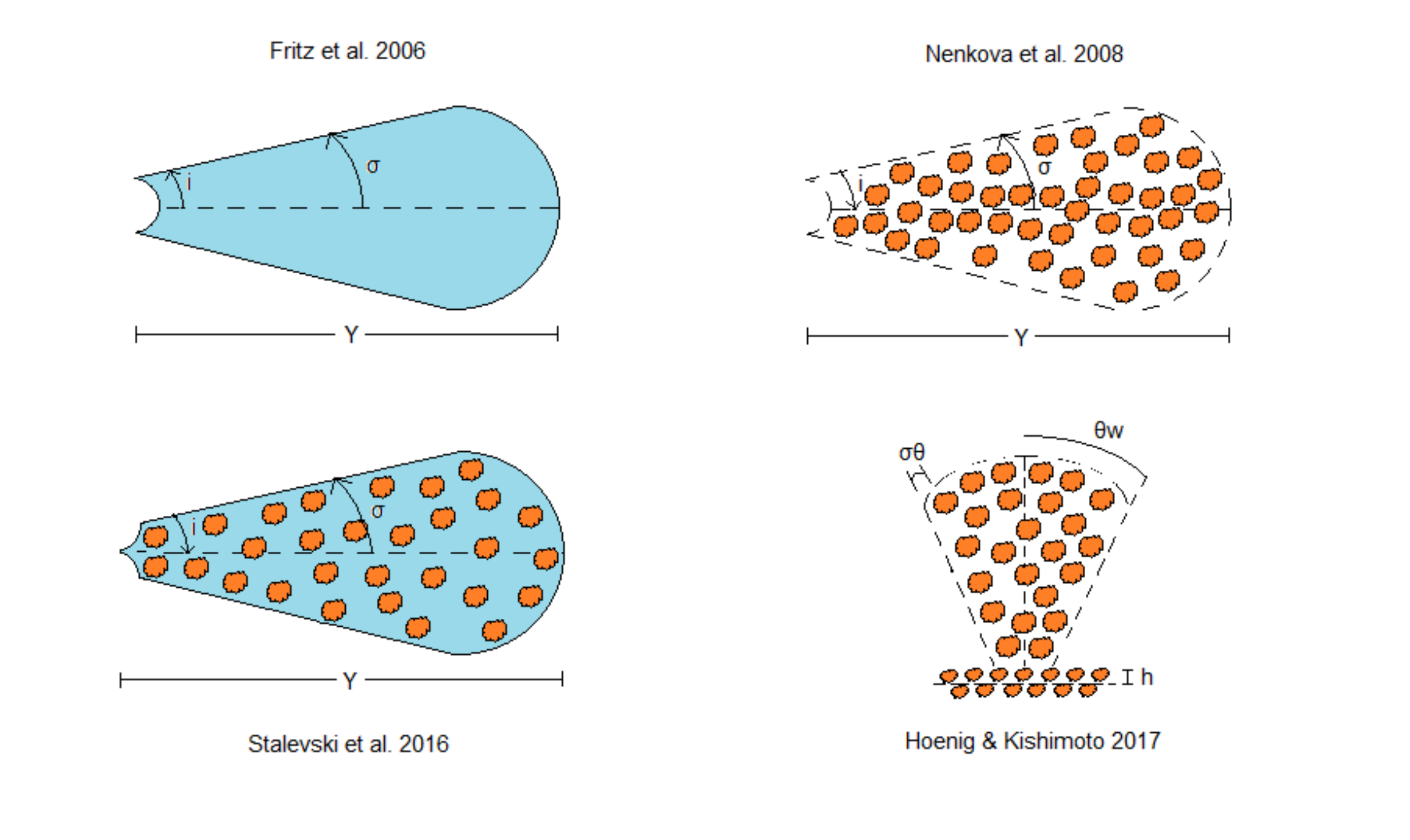}
\caption{Illustration of the dust models described in Sec.\,\ref{sec:Torus_models}: the smooth torus model by \cite{Fritz06} (\textit{top left}), the clumpy torus model by \cite{Nenkova08B} (\textit{top rigth}), the two phase torus model by \cite{Stalevski16} (\textit{bottom left}), and the clumpy disk and outflow model by \cite{Hoenig17} (\textit{bottom rigth}).}
\label{fig:models}
\end{center}
\end{figure*}

In order to show the geometry and dust distribution of the models tested in Sec.\,\ref{sec:Torus_models}, an sketch of the models is shown in figure\,\ref{fig:models}. The smooth torus model by \cite{Fritz06}, the clumpy torus model by \cite{Nenkova08B}, and two phase torus model by \cite{Stalevski16}, assume the dust distribution in a toroidal geometry. The clumpy disk and outflow model by \cite{Hoenig17} consider the dust distributed in a disk-like geometry plus a pollar hollow cone. The smooth torus model by \cite{Fritz06} considers a continuous distribution of dust toward the torus. The clumpy torus model by \cite{Nenkova08B} considers a toroidal distribution of dusty clumps. The two phase torus model by \cite{Stalevski16} considers a dust distribution of clumps embedded in a smooth component. The clumpy disk and outflow model by \cite{Hoenig17} considers a dust distribution in a clumpy disk plus a polar hollow cone.

\section{Models parameters} \label{sec:tables}

In order to reproduce the SED of NGC\,1068, we test the four dust models described in Section\,\ref{sec:Torus_models}. Our best fit assuming that a single SED of the models is able to reproduce N- and Q-bands at the same time is shown in Tables\,\ref{tab:Fritz_results}-\ref{tab:Hoenig_results} (Col.\,2). In all cases, many of the parameters are not well restricted. We also show the goodness of the fit throughout the $\chi^2$/dof.
Tables\,\ref{tab:Fritz_results}-\ref{tab:Hoenig_results} (Col.\,3 and onward) show the best fits obtained by unlinking one paremeter in Q-band. The smooth torus model by \cite{Fritz06} restricts all parameters by untying the optical depth at $\rm{9.7\mu m}$, $\rm{\tau_{9.7\mu m}}$. The clumpy torus model by \cite{Nenkova08B}, the two phase model by \cite{Stalevski16} and the clumpy disk and outflow model by \cite{Hoenig17} do not restrict all the parameters well in any case.
Finally, table\,\ref{tab:Stalevski_unlink_sigma}, shows the best fits for the two phase model by \cite{Stalevski16} untying the half opening angle with the other parameters. The table with the results for the smooth torus model by \cite{Fritz06} untying two parameters is shown in section\,\ref{sec:complex-SED}.\\
\\


\begin{table*}[ht!]
\scriptsize 
\begin{center}
\begin{tabular}{cc|ccccccc}\hline
\multicolumn{9}{c}{Smooth torus model by \cite{Fritz06}}\\\hline
\multirow{2}{*}{Param} & linking & \multirow{2}{*}{band} & \multicolumn{6}{c}{unlinking}\\
& parameters &  & $i$ & $\sigma$ & $\gamma$ & $\beta$ & $Y$ & $\tau$\\
\hline\hline
\multirow{2}{*}{i} & \multirow{2}{*}{$>0.01$} & N & $22.5\pm0.2$ & $60.16\pm^{1.07}_{1.32}$ & $49.15\pm^{3}_{10.1}$ & $>0.01$ & $34.76\pm^{0.6}_{1.5}$ & $1.13\pm^{0.58}_{0.48}$\\
& & Q & $>0.01$ & - & - & - & - & -\\
\multirow{2}{*}{$\sigma$} & \multirow{2}{*}{$>20$} & N & $>20$ & $>20$ & $>20$ & $>20$ & $23.94\pm^{2.33}_{0.96}$ & $57.63\pm^{0.8}_{2.23}$\\
& & Q & - & $58.5\pm^{00.7}_{0.42}$ & - & - & - & -\\
\multirow{2}{*}{$\gamma$} & \multirow{2}{*}{$1.998\pm^{0.004}_{0.006}$} & N & $3.99\pm0.004$ & $>0.01$ & $>0.01$ & $1.99\pm^{0.04}_{0.16}$ & $2.0\pm^{0.01}_{0.06}$ & $2.0\pm^{0.02}_{0.09}$\\
& & Q & - & - & ${5.42\pm^{0.04}_{0.12}}$ & - & - & -\\
\multirow{2}{*}{$\beta$} & \multirow{2}{*}{$-1.0$} & N & $-1.0$ & $-1.0$ & $-1.0$ & $-0.01$ & $-1.0$ & $-0.78\pm0.01$\\
& & Q & - & - & - & $-0.9\pm0.02$ & - & - \\
\multirow{2}{*}{Y} & \multirow{2}{*}{$>10$} & N & $>10$ & $>10$ & $59.93\pm^{1.53}_{1.28}$ & $>10$ & $>10$ & $12.64\pm^{0.27}_{0.31}$\\
& & Q & - & - & - & - & $>10$ & -\\
\multirow{2}{*}{$\rm{\tau_{9.7\mu m}}$} & \multirow{2}{*}{$2.82\pm^{0.03}_{0.01}$} & N & $<10$ & $1.99\pm^{0.03}_{0.05}$ & $1.73\pm^{0.04}_{0.02}$ & $2.0$ & $5.28\pm^{0.1}_{1.85}$ & $2.0\pm^{0.02}_{0.01}$\\
& & Q & - & - & - & - & - & $0.3\pm0.01$\\ \hline
$\chi^{2}/ dof$ & 1293.66/260 &  & 560.78/259 & 95.7/259 & 106.7/259 & 138.26/259 & 96.67/259 & 207.13/259\\
\hline
\end{tabular}
\end{center}
\caption{Fitting results of the N- and Q-band spectra to the smooth torus model by \citet{Fritz06}. Column\,(1): symbol of the parameter (see Table 1). Column\,(2): using a single SED by linking for the parameters in the N- and Q-band. Columns\,(3-9): unlinking one of the parameters in the Q-band with respect to those of the N-band. The $\rm{\chi^{2}/ dof}$ value for each fit is shown in the bottom row.}
\label{tab:Fritz_results}
\end{table*}


\begin{table*}[ht!]
\scriptsize 
\begin{center}
\begin{tabular}{cc|ccccccc}\hline
\multicolumn{9}{c}{Clumpy torus model by \cite{Nenkova08B}}\\\hline
\multirow{2}{*}{Param} & linking & \multirow{2}{*}{band} & \multicolumn{6}{c}{unlinking}\\
& parameters & & $i$ & $N0$ & $\sigma$ & $Y$ & $q$ & $\tau$\\
\hline\hline
\multirow{2}{*}{i} & \multirow{2}{*}{$>0.01$} & N & $>0.01$ & $>0.01$ & $>0.01$ & $>0.01$ & $>0.01$ & $>0.01$\\
& & Q & $>0.01$ & - & - & - & - & - \\
\multirow{2}{*}{N0} & \multirow{2}{*}{$<15$} & N & $<15$ & $<15$ & $<15$ & $<15$ & $<15$ & $<15$\\
&& Q & - & $5.87\pm^{0.08}_{0.03}$ & - & - & - & - \\
\multirow{2}{*}{$\sigma$} & \multirow{2}{*}{$<70$} & N & $56.44\pm^{0.25}_{0.78}$ & $65.07\pm^{0.11}_{0.54}$ & $64.86\pm^{0.23}_{0.67}$ & $<70$ & $<70$ & $64.92\pm^{0.2}_{0.22}$\\
& & Q & - & - & ${32.5\pm^{0.2}_{0.4}}$ & - & - & - \\ 
\multirow{2}{*}{Y} & \multirow{2}{*}{$<100$} & N & $<100$ & $<100$ & $<100$ & $<100$ & $<100$ & $<100$\\
&& Q & - & - & - & $<100$ & - & - \\
\multirow{2}{*}{q} & \multirow{2}{*}{$1.84\pm^{0.04}_{0.03}$} & N & $0.52\pm^{0.07}_{0.03}$ & $0.57\pm^{0.11}_{0.04}$ & $0.58\pm^{0.05}_{0.06}$ & $1.83\pm^{0.04}_{0.03}$ & $0.5\pm^{0.01}_{0.04}$ & $1.32\pm0.04$\\
&& Q & - & - & - & - & $<2.5$ & - \\
\multirow{2}{*}{$\tau_V$} & \multirow{2}{*}{$>10$} & N & $20.0\pm0.02$ & $20.0\pm^{0.01}_{0.05}$ & $19.99\pm0.3$ & $>10$ & $13.48\pm^{0.19}_{0.18}$ & $20.0\pm0.02$\\
& & Q & - & - & - & - & - & $>10$\\ \hline
$\chi^{2}/ dof$ & 2682.42/260 & & 1655.51/259 & 1496.7/259 & 1476.82/259 & 2682.42/259 & 1913.35/259 & 1722.95/259\\
\hline
\end{tabular}
\end{center}
\caption{Same as in table\,\ref{tab:Fritz_results} for the clumpy torus model by \citet{Nenkova08B}.}
\label{tab:Nenkova_results}
\end{table*}


\begin{table*}[ht!]
\scriptsize 
\begin{center}
\begin{tabular}{cc|ccccccc}\hline
\multicolumn{9}{c}{Two phase torus model by \cite{Stalevski16}}\\\hline
\multirow{2}{*}{Param} & linking & \multirow{2}{*}{band} & \multicolumn{6}{c}{unlinking}\\
& parameters & & $i$ & $\sigma$ & $p$ & $q$ & $Y$ & $\tau$\\
\hline\hline
\multirow{2}{*}{$i$} & \multirow{2}{*}{$>0.01$} & N & $40.0\pm^{0.05}_{0.08}$ & $88.66\pm^{0.23}_{0.9}$ & $>0.01$ & $>0.01$ & $88.84\pm0.02$ & $89.15\pm0.02$\\
&& Q & $>0.0190$ & - & - & - & - & - \\

\multirow{2}{*}{$\sigma$} & \multirow{2}{*}{$69.9\pm^{0.04}_{0.44}$} & N & $69.99\pm0.05$ & $11.69\pm^{0.07}_{0.04}$ & $69.99\pm^{0.04}_{0.73}$ & $53.59\pm^{0.44}_{0.63}$ & $<80$ & $<80$\\
&& Q & - & $<80$ & - & - & - & - \\

\multirow{2}{*}{$p$} & \multirow{2}{*}{$<1.5$} & N & $<1.5$ & $<1.5$ & $<1.5$ & $<1.5$ & $<1.5$ & $<1.5$\\
&& Q & - & - & $<1.5$ & - & - & - \\

\multirow{2}{*}{$q$} & \multirow{2}{*}{$<1.5$} & N & $<1.5$ & $0.54\pm^{0.46}_{0.28}$ & $<1.5$ & $<1.5$ & $<1.5$ & $<1.5$\\
&& Q & - & - & - & $<1.5$ & - & - \\

\multirow{2}{*}{$Y$} & \multirow{2}{*}{$<30$} & N & $<30$ & $<30$ & $<30$ & $<30$ & $<30$ & $<30$\\
&& Q & - & - & - & - & $<30$ & - \\
\multirow{2}{*}{$\rm{\tau_{9.7\mu m}}$} & \multirow{2}{*}{$5.0$} & N & $<11$ & $<11$ & $5.0$ & $<11$ & $<11$ & $<11$\\
&& Q & - & - & - & - & - & $7.46\pm0.07$ \\ \hline
$\chi^{2}/ dof$ & $3021.44/260$ & & $2390.52/259$ & $205.26/259$ & $3021.32/259$ & $1256.69/259$ & $1042.39/259$ & $677.04/259$\\
\hline
\end{tabular}
\end{center}
\caption{Same as in table\,\ref{tab:Fritz_results} for the two phase torus model by \cite{Stalevski16}.}
\label{tab:Stalevski_results}
\end{table*}


\begin{table*}[ht!]
\scriptsize 
\begin{center}
\begin{tabular}{cc|ccccccccc}\hline
\multicolumn{11}{c}{Clumpy disk and outflow model by \cite{Hoenig17}}\\\hline
\multirow{2}{*}{Param} & linking & \multirow{2}{*}{band} & \multicolumn{8}{c}{unlinking}\\
& parameters & & $i$ & $N0$ & $a$ & $\theta$ & $\sigma$ & $a_w$ & $h$ & $f_{wd}$\\
\hline\hline

\multirow{2}{*}{i} & \multirow{2}{*}{$>0.01$} & N & $>0.01$ & $>0.01$ & $>0.01$ & $>0.01$ & $>0.01$ & $>0.01$ & $>0.01$ & $>0.01$ \\
&& Q & $>0.01$ & - & - & - & - & - & - & - \\

\multirow{2}{*}{N0} & \multirow{2}{*}{$<10$} & N & $<10$ & $7.0\pm0.01$ & $5.99\pm0.1$ & $7.0\pm^{0.01}_{0.03}$ & $7.0$ & $<10$ & $5.44\pm^{0.04}_{0.05}$ & $7.0$ \\
&& Q & - & $<10$ & - & - & - & - & - & - \\

\multirow{2}{*}{a} & \multirow{2}{*}{$-2.15\pm0.01$} & N & $-2.15\pm0.01$ & $-2.07$ & $-2.0$ & $-2.04$ & $-2.0$ & $-2.0$ & $-2.0$ & $-2.0$ \\
&& Q & - & - & $-2.45\pm^{0.02}_{0.01}$ & - & - & - & - & - \\

\multirow{2}{*}{$\theta$} & \multirow{2}{*}{$<15$} & N & $<15$ & $<15$ & $<15$ & $<15$ & $<15$ & $10.0\pm^{0.03}_{0.01}$ & $10.0\pm^{0.04}_{0.01}$ & $<15$\\
&& Q & - & - & - & $<15$ & - & - & - & - \\

\multirow{2}{*}{$\sigma$} & \multirow{2}{*}{$<45$} & N & $<45$ & $<45$ & $<45$ & $<45$ & $<45$ & $<45$ & $<45$ & $<45$\\
&& Q & - & - & - & - & $<45$ & - & - & - \\

\multirow{2}{*}{$a_{w}$} & \multirow{2}{*}{$>-2.5$} & N & $>-2.5$ & $>-2.5$ & $>-2.5$ & $>-2.5$ & $>-2.5$ & $-2.38\pm0.01$ & $>-2.5$ & $>-2.5$\\
&& Q & - & - & - & - & - & $>-2.5$ & - & - \\

\multirow{2}{*}{h} & \multirow{2}{*}{$<0.5$} & N & $<0.5$ & $<0.5$ & $<0.5$ & $<0.5$ & $<0.5$ & $<0.5$ & $0.14$ & $<0.5$\\
&& Q & - & - & - & - & - & - & $<0.5$ & - \\

\multirow{2}{*}{$f_{wd}$} & \multirow{2}{*}{$0.6$} & N & $0.6$ & $0.6$ & $0.6$ & $<0.75$ & $0.71$ & $0.6$ & $0.6$ & $0.69\pm0.01$\\
&& Q & - & - & - & - & - & - & - & $<0.75$ \\ \hline

$\chi^{2}/ dof$ & 1138.56/258 & & 1138.51/257 & 651.11/257 & 618.99/257 & 736.4/257 & 540.26/257 & 1057.5/257 & 769.98/257 & 538.73/257\\
\hline
\end{tabular}
\end{center}
\caption{Same as in table\,\ref{tab:Fritz_results} for the Clumpy disk plus outflow model by \cite{Hoenig17}.}
\label{tab:Hoenig_results}
\end{table*}


\clearpage


\begin{table}[ht!]
\scriptsize 
\begin{center}
\begin{tabular}{ccccccc}\hline
\multicolumn{6}{c}{Two phase torus model by \cite{Stalevski16}}\\\hline
\multirow{2}{*}{Param} & \multirow{2}{*}{band} & \multicolumn{4}{c}{unlinking}\\
& & $\sigma/p$ & $\sigma/q$ & $\sigma/Y$ & $\sigma/\tau$\\
\hline\hline
\multirow{2}{*}{$i$} & N & $80.0\pm^{0.1}_{0.03}$ & $80.0\pm^{0.13}_{0.05}$ & $89.17\pm^{0.03}_{0.07}$ & $80.0\pm^{0.11}_{0.08}$\\
& Q & - & - & - & - \\

\multirow{2}{*}{$\sigma$} & N & $19.0\pm^{1.35}_{0.85}$ & $19.25\pm^{0.28}_{0.16}$ & $10.67\pm^{0.08}_{0.03}$ & $18.53\pm^{0.45}_{0.13}$\\
& Q & $11.8$ & ${11.99\pm^{0.16}_{0.09}}$ & $10.0$ & ${11.59\pm0.1}$\\

\multirow{2}{*}{$p$} & N & $<1.5$ & $<1.5$ & $<1.5$ & $<1.5$\\
& Q & $<1.5$ & - & - & -\\

\multirow{2}{*}{$q$} & N & $<1.5$ & $<1.5$ & $<1.2$ & $<1.5$\\
& Q & - & $<1.5$ & - & -\\

\multirow{2}{*}{$Y$} & N & $<30$ & $<30$ & $<30$ & $<30$\\
& Q & - & - & $<30$ & -\\

\multirow{2}{*}{$\tau$} & N & $3.93\pm^{0.11}_{0.04}$ & $3.9\pm^{0.13}_{0.04}$ & $<11$ & $4.04\pm^{0.14}_{0.04}$\\
& Q & - & - & - & $<11$\\ \hline
$\chi^{2}/ dof$ & & $207.14/258$ & $184.63/258$ & $145.01/258$ & $174.9/258$\\
\hline
\end{tabular}
\end{center}
\caption{Same as in table\,\ref{tab:Fritz_unlink_two} for the two phase torus model by \cite{Stalevski16} when we unlink two parameters between N- and Q-bands. }
\label{tab:Stalevski_unlink_sigma}
\end{table}

\section{The bayes factor} \label{sec:AIC}


In order to evaluate to what extent a model is better than another one we calculate the bayes factor through the Akaike information criterion, $\rm{AIC_c}$. To this end, we use the eq.\,5 in \cite{Emmanoulopoulos16} to calculate the $\rm{AIC_c}$

\begin{equation}
    AIC_c = 2k - 2C_L + \chi^2 + \frac{2k(k+1)}{N-k-1}
\end{equation}

\noindent where $C_L$ is the constant likelihood of the true hypothetical model, $k$ is the number of free model parameters, and $N$ is the number of data points.

We then calculate the difference between two different models, $\rm{\Delta[AIC_{c}]}$

\begin{equation}
    \Delta[AIC_{c}] = AIC_{c,2} - AIC_{c,1}
\end{equation}

Finally, we estimate the evidence ratio, $\rm{\epsilon}$ 

\begin{equation}
    \epsilon = e^-\frac{\Delta[AIC_{c}]}{2}
\end{equation}


\noindent The evidence ratio or bayes factor, is a measure of the relative likelihood of one versus other model. When the bayes factor is $\rm{\leq 0.01}$, the first model is more likely to be the `correct' model. When the bayes factor is $\rm{\geq 100}$ the second model is  more likely to be the `correct' model.

\section{Graphite/silicate fraction} \label{sec:proportion}

\begin{figure}[ht!]
\begin{center}
\includegraphics[width=0.5\columnwidth]{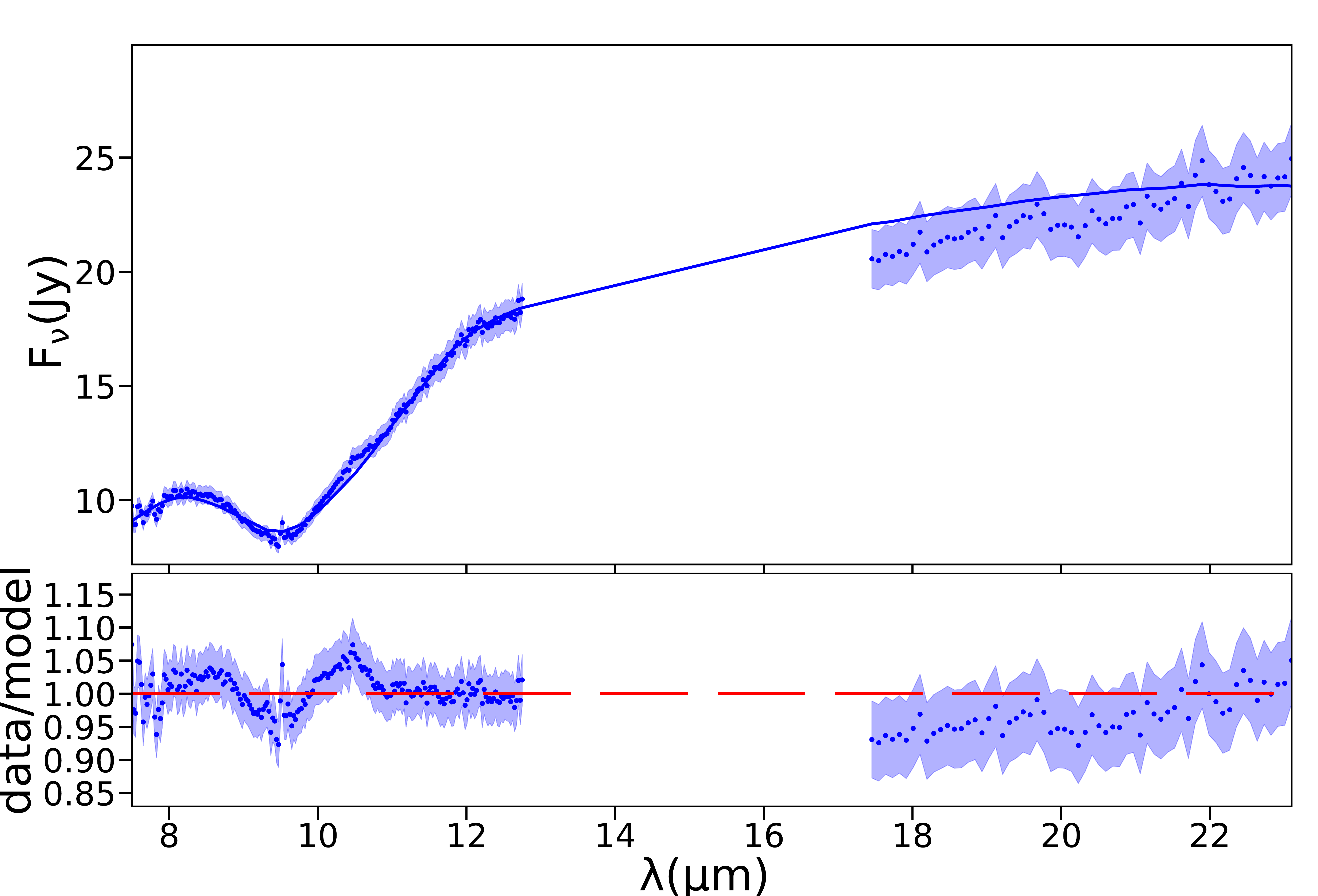}
\caption{Best model exploring for different graphite/silicate ratio (see Section\,\ref{sec:proportion}). The description in this figure is the same as that reported in Fig.\,\ref{fig:first_second_angle}.}
\label{fig:ratio}
\end{center}
\end{figure}

Several of the AGN dust models \citep[e.g.][]{Hoenig10B} rely on composition constants derived by \citet{Mathis77} for the ISM. In particular, the normalization with respect to hydrogen abundance are log(A) = $\rm{-15.24}$ and $\rm{-15.21}$ for graphite and silicate grains, respectively (i.e. 51.7\% of silicate). However, \citet{Draine-Lee84} updated these numbers to log(A) = $\rm{-25.11}$ and $\rm{-25.16}$ for silicate and graphite, respectively (i.e. 52.9\% of silicate). This is the composition assumed by the smooth torus model by \citet{Fritz06}. \citet{Weingartner01} also gave the same abundance for the silicate but different number for graphite with log(A) = $\rm{-25.13}$ (i.e. 52.4\% of silicate). This is the assumed normalization factors in the two-phase torus model presented by \citet{Stalevski16}. This fraction might have an impact on the results. 
In Section\,\ref{sec:Spectral_fitting} we used a ratio of 49\% and 51\% for graphite and silicate grains, respectively. However, this is further explored in this section. We create a grid covering a range percentages of silicates, $\rm{f_{sil}}$ = [30,40,45-55,60 and 70]\%. Notice that in this grid, we do not include the previously used ratio of particles (49\% and 51\% for graphite and silicate grains, respectively). We obtained the best fit with $\rm{\chi_{r}^{2} = 0.51}$ for 51\% graphite and 49\% silicates. This fit has a bayes factor of $4 \times 10^{-7}$ compared to the best fit obtained in Section\,\ref{sec:tau}. Fig.\,\ref{fig:ratio} shows the best fit obtained with 51\% graphite and 49\% silicate.

\begin{figure}[ht!]
\begin{center}
\includegraphics[width=0.45\columnwidth]{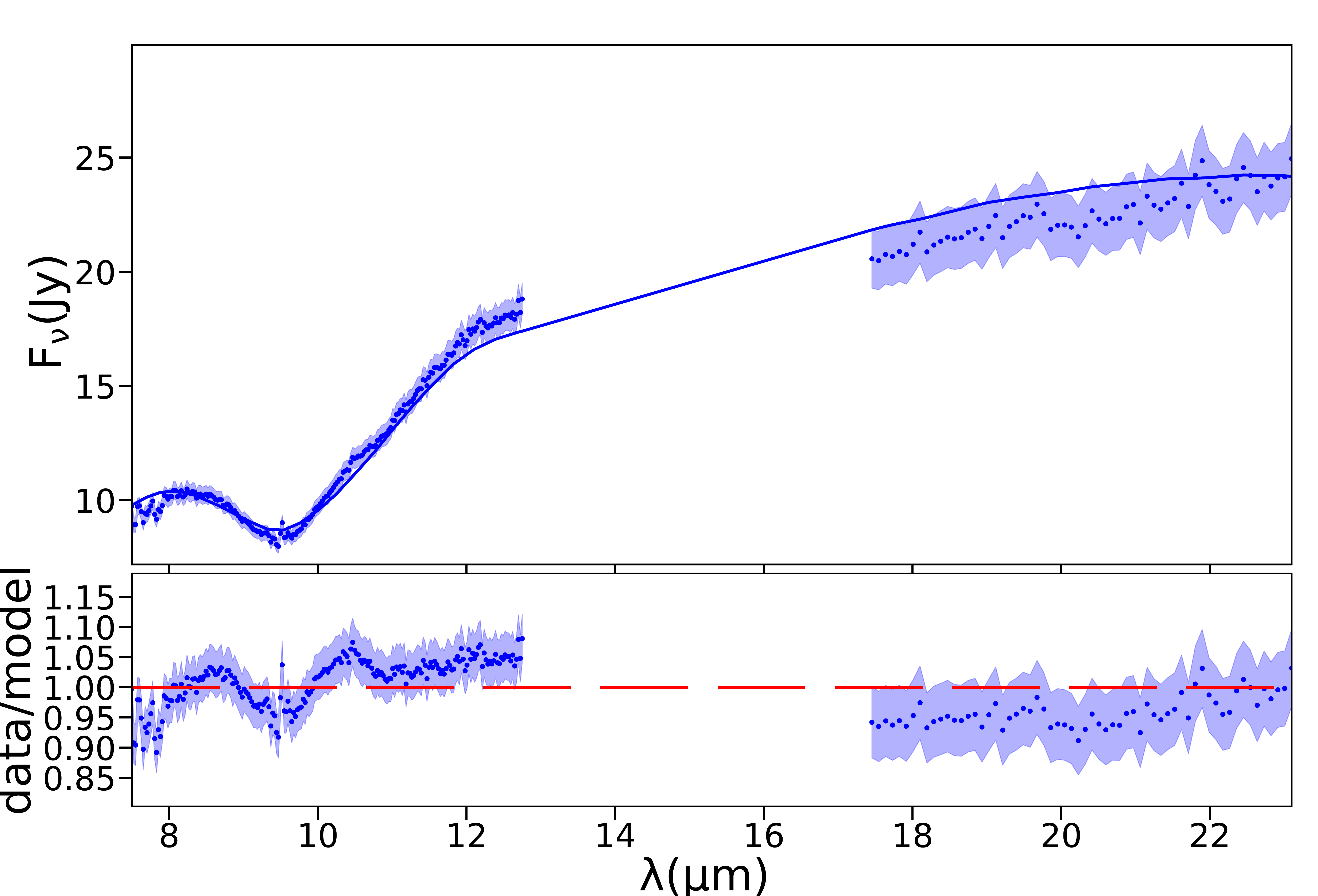}
\includegraphics[width=0.45\columnwidth]{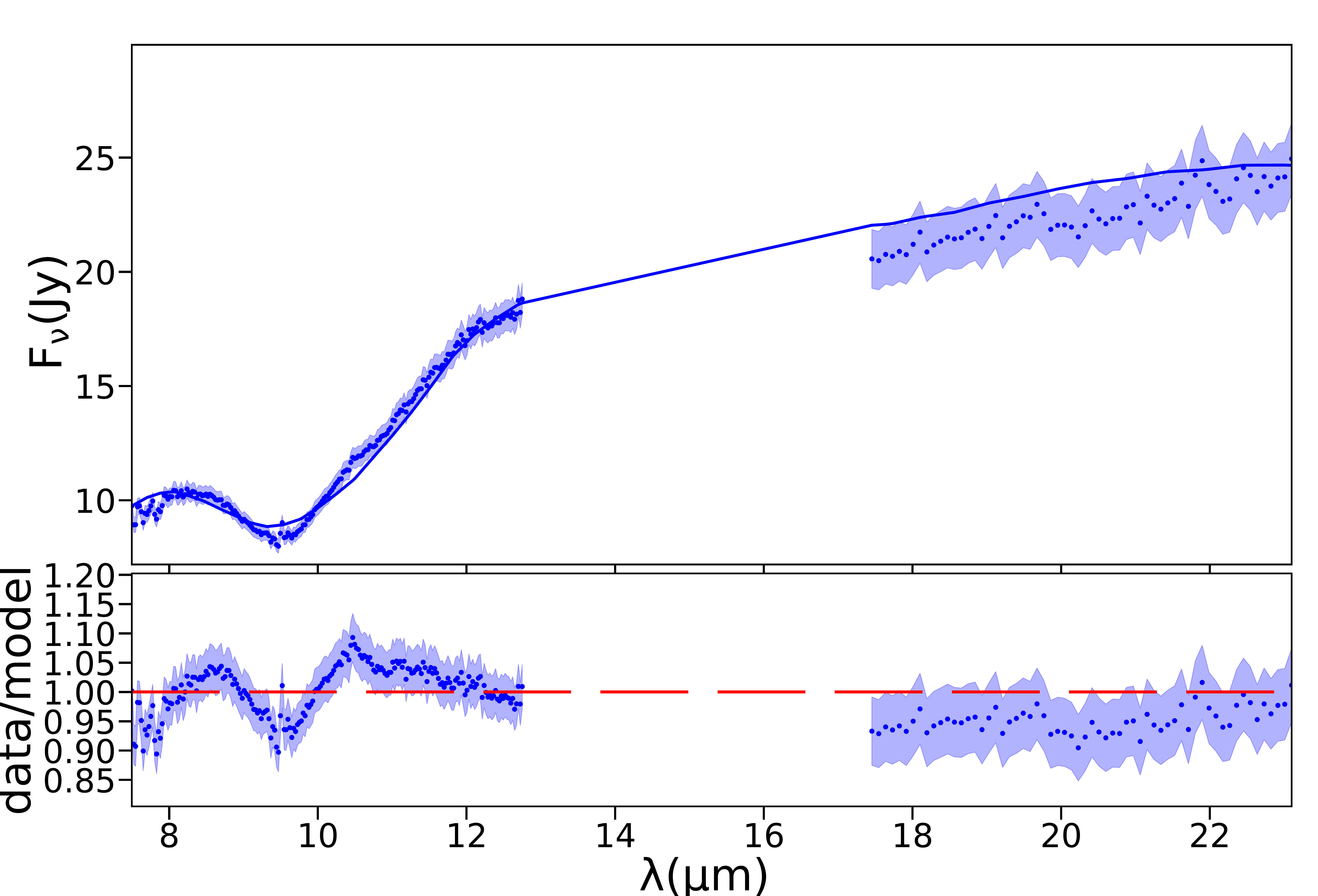}\\
\includegraphics[width=0.45\columnwidth]{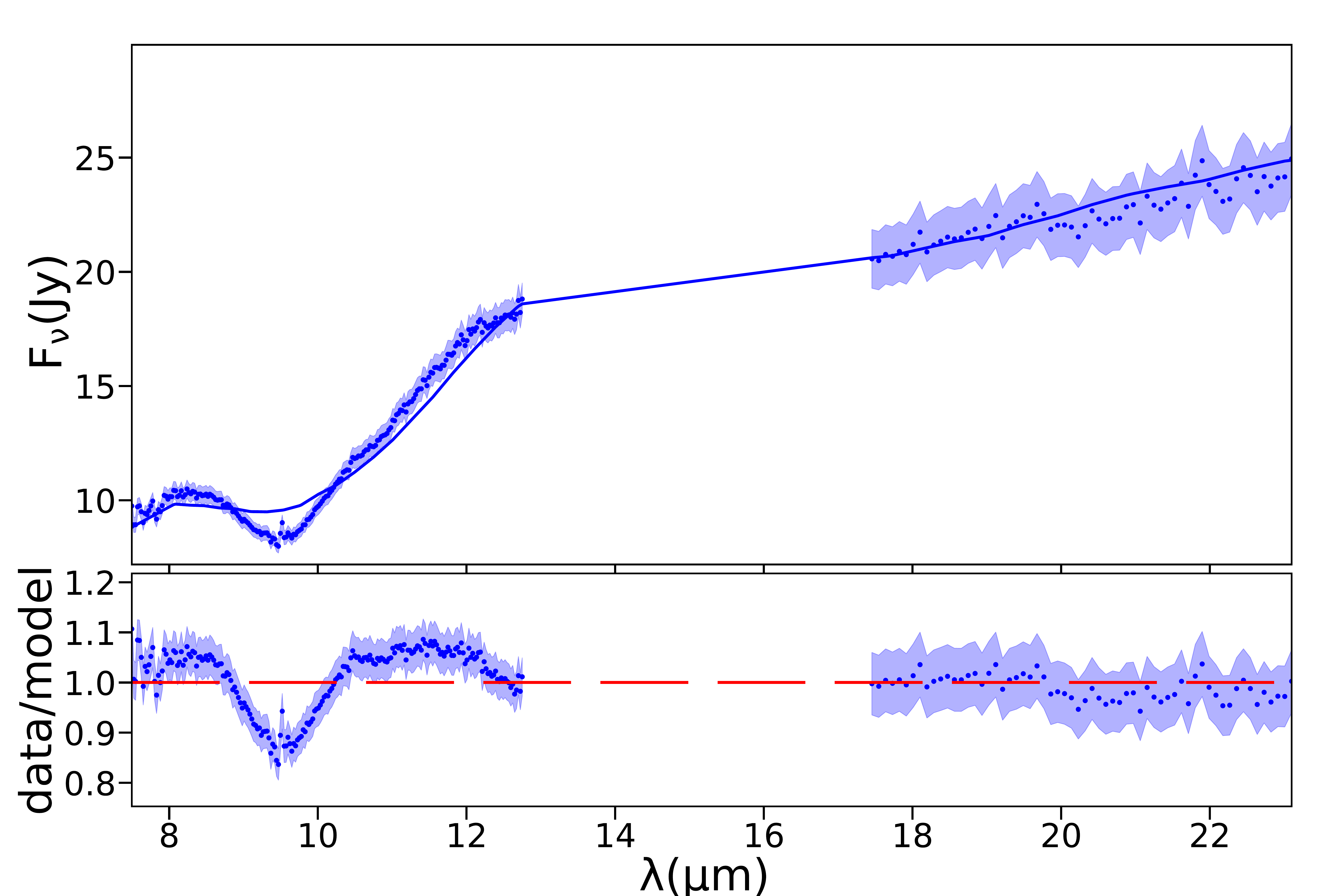}
\includegraphics[width=0.45\columnwidth]{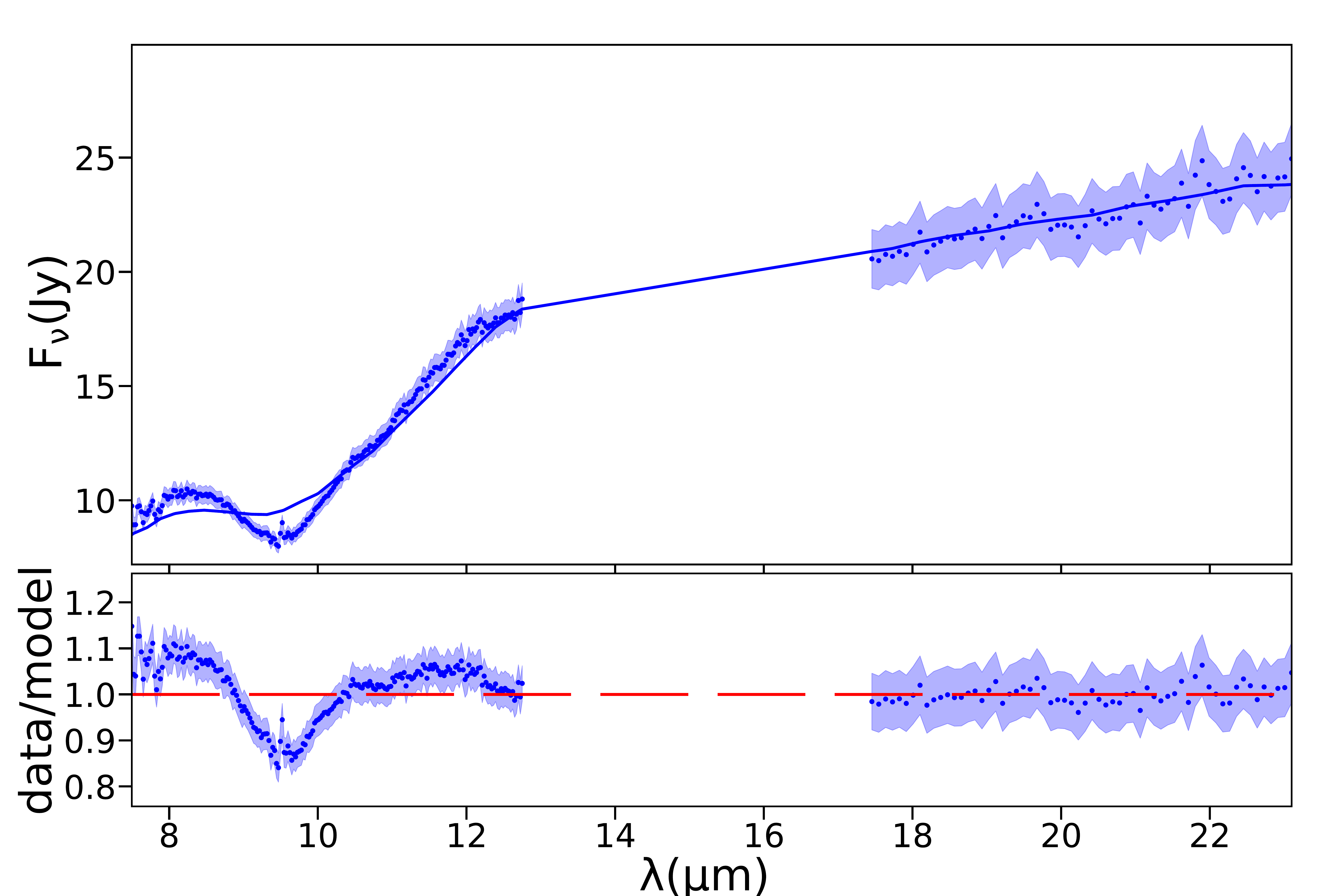}
\caption{(top-left): Best model obtained when we test for different inner radius for the largest torus. (top-right): Best fit obtained from model with reshape inner wall of the torus. (bottom): Best fit when we consider different sublimation temperature for graphite and silicate using T(Graphite)=1500\,K and T(Silicate)=1000\,K (left) and T(Graphite)=2000 K and T(Silicate)=1500\,K (right).  The description in this figure is the same as that reported in Fig.\,\ref{fig:first_second_angle}.} 
\label{fig:infructuosos}
\end{center}
\end{figure}

\section{Inner radius of the torus} \label{sec:4_torus}

We explored the scenario when the inner radius of the largest torus is different to that of the smallest torus. We initially kept the inner radius of the smallest torus fixed to 0.2\,pc and we tested a range of values of the largest torus: $\rm{R_{in,2} = [0.1,0.2,0.3,0.4,0.5,0.6]}$\,pc. The other parameters are fixed to the best values obtained in section\,\ref{sec:tau}. The grid has 2376 SEDs and the best fit was obtained with $\rm{\chi_{r}^{2} = 1.06}$ for $\rm{R_{in,2}=0.1}$\,pc, with inclination angle of $\rm{i=61^\circ}$ and $\rm{A_v=5}$ magnitudes. This fit has a bayes factor of $3.9 \times 10^{-39}$ compared to the best fit in Section\,\ref{sec:tau}. Fig.\,\ref{fig:infructuosos} (top left panel) shows this result. 

Graphite grains can sustain higher temperatures than silicate grains, with the former being able to heat up to $\sim$ 1900-2000\,K and the latter sublimating at $\sim$ 800-1200 K, depending on density \citep[][and references therein]{Garcia-Gonzalez17}. This implies a different sublimation radius for the graphite compared to silicate grains. In order to explore this, we create a complex system of tori, composed by two tori for graphite and two tori for silicate grains. All the parameters are the same, changing the sublimation radius for each kind of particle. For this, we made two grids of SEDs, first we assume a sublimation temperature of 1500 and 1000\,K for graphite and silicate respectively (i.e. the sublimation radii for graphite and silicate grains at 0.4 and 1.2\,pc, respectively). However, this fit is not satisfactory ($\rm{\chi_{r}^{2} = 2.38}$), with $i=64^\circ$ and $\rm{A_V = 0}$ mag. Then, we assume a sublimation temperature of 2000 and 1500\,K for graphite and silicate, respectively (i.e. the sublimation radius for graphite and silicate was set to 0.17 and 0.4\,pc, respectively). However, this fit is not statistically acceptable with $\rm{\chi_{r}^{2} = 2.48}$, with $i\rm{=62^\circ}$ and $\rm{A_V = 0}$ mag. These SEDs has a bayes factor of $5.9 \times 10^{-116}$ and $9 \times 10^{-122}$ compared to the best fit in Section\,\ref{sec:tau}, respectively. Fig.\,\ref{fig:infructuosos} (bottom panels) shows these two fits. While they are quite good at reproducing the slope between the N and Q bands, they fail to reproduce the 10$\rm{\mu m}$ silicate feature.  

Finally, the inner region of the torus might be reshaped to account for this anisotropic irradiation by the accretion disk, with the strongest emission perpendicular to the disk and none in the equatorial plane \citep{Stalevski16}. In this case, the incident flux is a function of the distance and the polar angle. In order to explore this, we used the anisotropic disk emission by \citet{Netzer87}, where the flux as a function of the polar angle $\Theta$ is (see also \citet{Netzer13}):

\begin{equation}
    F_{AGN} = (1/3)cos\theta(1+2cos\theta)
\end{equation}

In this case, the sublimation radius is:

\begin{equation}
    R_{d}(\theta) = (0.4)L_{45}(\theta)^{1/2}Tsub^{-2.6}_{1500}pc
\end{equation}

\noindent where $\rm{L_{45}(\theta) = [s+(1-s)(1/3)cos\theta(1+2cos\theta)]L_{bol_{45}}}$, where $\rm{L_{bol_{45}}}$ is the AGN luminosity in units of $\rm{10^{45}erg/s}$ and $s$ is the softening factor introduced to prevent the inner dust radius to reach zero at the equatorial plane ($s$=1 recovers the isotropic scenario). 

We created a new grid including 48 SEDs with a range of inner radii of $\rm{R_{in}}=[0.2,0.3,0.4,0.5,0.6,0.7,1.0]$\,pc, and a cut-off radius at half of the $\rm{R_{in}}$ (i.e. $\rm{R_{cutoff}}$ = [0.1, 0.15, 0.2, 0.25, 0.3, 0.35, 0.5]\,pc). We have obtained that the best fit shows $\rm{\chi_{r}^{2} = 1.14}$ for $\rm{R_{in}}$=0.5 pc and $\rm{R_{cutoff}}$=0.25\,pc, with inclination angle of $\rm{68^\circ}$ and extinction of 0 magnitudes (see Fig.\,\ref{fig:infructuosos} top right panel). The reshape of the inner radius of the torus is not an improvement compared to those SEDs obtained without taking this effect into account. Indeed, the model shown in Section\,\ref{sec:tau} without reshape has a bayes factor of $1.1 \times 10^{43}$ compared to this fit. 

\section{Non-concentric tori} \label{sec:non-concentric tori}

\begin{figure}[ht!]
\begin{center}
\includegraphics[width=0.5\columnwidth]{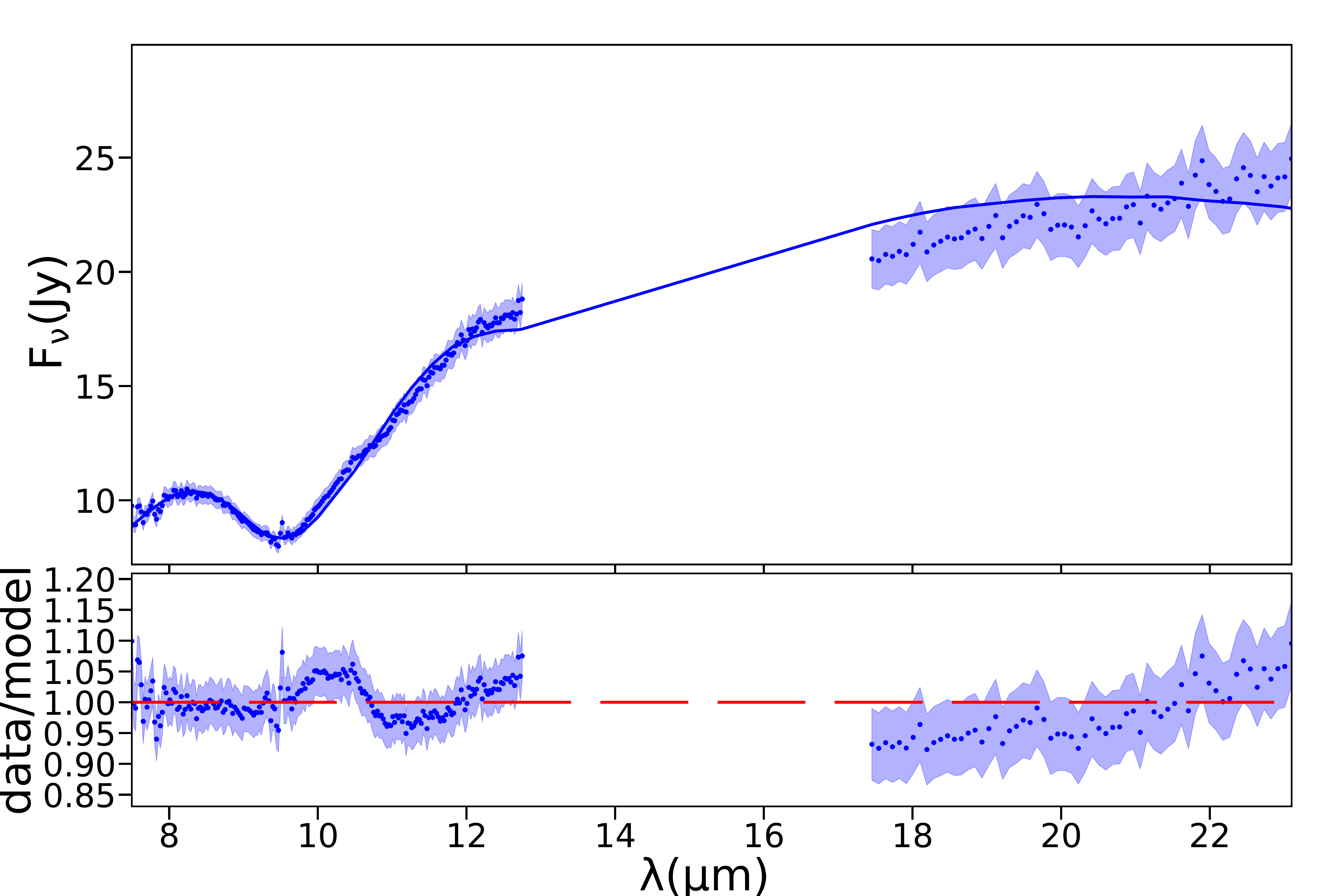}
\caption{Best model obtained when we test for two non-concentric tori (see Section.\,\ref{sec:non-concentric tori}). The description in this figure is the same as that reported in Fig.\,\ref{fig:first_second_angle}.}
\label{fig:offsetX}
\end{center}
\end{figure}

In order to test the scenario in which the two components found through our analysis are associated to two different AGN (according to recent literature, see Discussion section), we produced new synthetic SEDs for two non-concentric tori, with the same parameters obtained in our final best model but considering each torus host an AGN. For simplicity, we assume that the luminosity of each AGN is half of the total luminosity of NGC\,1068. We explored several distances between the two AGN, from 0.1 to 25 pc (11 steps), where the radiation from both AGN reaches both tori, and from 30 to 40 pc (6 steps), where each AGN affects only one of the tori. We also explore the resulting SED when the separation occurs along the height or along the width of the torus. Our best model shows a statistic of $\rm{\chi_{r}^{2} = 0.63}$ with an offset of 10\,pc between each AGN, a viewing angle of $\rm{90^\circ}$ and an foreground extinction of $\rm{A_V=3}$ mag. This fit has a bayes factor of $4.1 \times 10^{-14}$ compared to the best fit in Section\,\ref{sec:tau}. Fig.\,\ref{fig:offsetX} shows the resulting SED with these two non-concentric tori. Although the fit is quite good, in particular around the silicate absorption feature at 10$\rm{\mu m}$, it is not better than that provided by the two concentric torus, discussed in the main body of this paper.

\clearpage
\begin{thebibliography}{}

\bibitem[Absil et al.(2013)]{Absil13} Absil, O., Defr{\`e}re, D., Coud{\'e} du Foresto, V., et al.\ 2013, \aap, 555, A104. doi:10.1051/0004-6361/201321673

\bibitem[Alonso-Herrero et al.(2011)]{Alonso-Herrero11} Alonso-Herrero, A., Ramos Almeida, C., Mason, R., et al.\ 2011, \apj, 736, 82 

\bibitem[Antonucci \& Miller(1985)]{Antonucci85} Antonucci, R.~R.~J. \& Miller, J.~S.\ 1985, \apj, 297, 621. doi:10.1086/163559

\bibitem[Antonucci(1993)]{Antonucci93} Antonucci, R.\ 1993, \araa, 31, 473 

\bibitem[Arnaud(1996)]{Arnaud96} Arnaud, K.~A.\ 1996, in ASP Conf. Ser. 101, Astronomical Data Analysis Software and Systems V, ed. G. H. Jacoby \& J. Barnes (San Francisco, CA: ASP), 17

\bibitem[Asmus(2019)]{Asmus19} Asmus, D.\ 2019, \mnras, 489, 2177. doi:10.1093/mnras/stz2289

\bibitem[Baes et al.(2003)]{Baes03} Baes, M., Davies, J.~I., Dejonghe, H., et al.\ 2003, \mnras, 343, 1081

\bibitem[Baes et al.(2011)]{Baes11} Baes, M., Verstappen, J., De Looze, I., et al.\ 2011, \apjs, 196, 22

\bibitem[Bauer et al.(2015)]{Bauer15} Bauer, F.~E., Ar{\'e}valo, P., Walton, D.~J., et al.\ 2015, \apj, 812, 116. doi:10.1088/0004-637X/812/2/116

\bibitem[Bland-Hawthorn et al.(1997)]{Bland-Hawthorn97} Bland-Hawthorn, J., Gallimore, J.~F., Tacconi, L.~J., et al.\ 1997, \apss, 248, 9. doi:10.1023/A:1000567831370

\bibitem[Bock et al.(2000)]{Bock00} Bock, J.~J., Neugebauer, G., Matthews, K., et al.\ 2000, \aj, 120, 2904. doi:10.1086/316871

\bibitem[Boley et al.(2013)]{Boley13} Boley, P.~A., Linz, H., van Boekel, R., et al.\ 2013, \aap, 558, A24.

\bibitem[Calvet et al.(2002)]{Calvet02} Calvet, N., D'Alessio, P., Hartmann, L., et al.\ 2002, \apj, 568, 1008. doi:10.1086/339061

\bibitem[Calzetti et al.(2000)]{Calzetti00} Calzetti, D., Armus, L., Bohlin, R.~C., et al.\ 2000, \apj, 533, 682

\bibitem[Cameron et al.(1993)]{Cameron93} Cameron, M., Storey, J.~W.~V., Rotaciuc, V., et al.\ 1993, \apj, 419, 136. doi:10.1086/173467

\bibitem[Das et al.(2006)]{Das06} Das, V., Crenshaw, D.~M., Kraemer, S.~B., et al.\ 2006, \aj, 132, 620. doi:10.1086/504899


\bibitem[Draine \& Lee(1984)]{Draine-Lee84} Draine, B.~T. \& Lee, H.~M.\ 1984, \apj, 285, 89. doi:10.1086/162480

\bibitem[Efstathiou, \& Rowan-Robinson(1995)]{Efstathiou95} Efstathiou, A., \& Rowan-Robinson, M.\ 1995, \mnras, 273, 649

\bibitem[Emmanoulopoulos et al.(2016)]{Emmanoulopoulos16} Emmanoulopoulos, D., Papadakis, I.~E., Epitropakis, A., et al.\ 2016, \mnras, 461, 1642

\bibitem[Esparza-Arredondo et al.(2019)]{Esparza-Arredondo19} Esparza-Arredondo, D., Gonz{\'a}lez-Mart{\'\i}n, O., Dultzin, D., et al.\ 2019, \apj, 886, 125. doi:10.3847/1538-4357/ab4ced


\bibitem[Esparza Arredondo et al.(2021)]{Esparza-Arredondo21} Esparza Arredondo, D., Gonz{\'a}lez Mart{\'\i}n, O., Dultzin, D., et al.\ 2021, arXiv:2104.11263

\bibitem[Feltre et al.(2012)]{Feltre12} Feltre, A., Hatziminaoglou, E., Fritz, J., et al.\ 2012, \mnras, 426, 120. doi:10.1111/j.1365-2966.2012.21695.x

\bibitem[Fritz et al.(2006)]{Fritz06} Fritz, J., Franceschini, A., \& Hatziminaoglou, E.\ 2006, \mnras, 366, 767

\bibitem[Gallimore et al.(2004)]{Gallimore2004} Gallimore, J.~F., Baum, S.~A., \& O'Dea, C.~P.\ 2004, \apj, 613, 794

\bibitem[Garc{\'\i}a-Bernete et al.(2019)]{Garcia-Bernete19} Garc{\'\i}a-Bernete, I., Ramos Almeida, C., Alonso-Herrero, A., et al.\ 2019, \mnras, 486, 4917. doi:10.1093/mnras/stz1003


\bibitem[Garc{\'\i}a-Burillo et al.(2014)]{Garcia-Burillo14} Garc{\'\i}a-Burillo, S., Combes, F., Usero, A., et al.\ 2014, \aap, 567, A125. doi:10.1051/0004-6361/201423843

\bibitem[Garc{\'{\i}}a-Burillo et al.(2016)]{Garcia-Burillo16} Garc{\'{\i}}a-Burillo, S., Combes, F., Ramos Almeida, C., et al.\ 2016, \apjl, 823, L12 

\bibitem[Garc{\'\i}a-Burillo et al.(2019)]{Garcia-Burillo19} Garc{\'\i}a-Burillo, S., Combes, F., Ramos Almeida, C., et al.\ 2019, \aap, 632, A61. doi:10.1051/0004-6361/201936606

\bibitem[Garc{\'\i}a-Burillo et al.(2021)]{Garcia-Burillo21} Garc{\'\i}a-Burillo, S., Alonso-Herrero, A., Ramos Almeida, C., et al.\ 2021, \aap, 652, A98. doi:10.1051/0004-6361/202141075


\bibitem[Garc{\'\i}a-Gonz{\'a}lez et al.(2017)]{Garcia-Gonzalez17} Garc{\'\i}a-Gonz{\'a}lez, J., Alonso-Herrero, A., H{\"o}nig, S.~F., et al.\ 2017, \mnras, 470, 2578

\bibitem[Gonz{\'a}lez-Mart{\'\i}n et al.(2013)]{Gonzalez-Martin13} Gonz{\'a}lez-Mart{\'\i}n, O., Rodr{\'\i}guez-Espinosa, J.~M., D{\'\i}az-Santos, T., et al.\ 2013, \aap, 553, A35

\bibitem[Gonz{\'a}lez-Mart{\'\i}n et al.(2019)]{Gonzalez-Martin19A} Gonz{\'a}lez-Mart{\'\i}n, O., Masegosa, J., Garc{\'\i}a-Bernete, I., et al.\ 2019, \apj, 884, 10. doi:10.3847/1538-4357/ab3e6b


\bibitem[Gonz{\'a}lez-Mart{\'\i}n et al.(2019)]{Gonzalez-Martin19B} Gonz{\'a}lez-Mart{\'\i}n, O., Masegosa, J., Garc{\'\i}a-Bernete, I., et al.\ 2019, \apj, 884, 11. doi:10.3847/1538-4357/ab3e4f

\bibitem[Goosmann \& Gaskell(2007)]{Goosmann07} Goosmann, R.~W. \& Gaskell, C.~M.\ 2007, \aap, 465, 129. doi:10.1051/0004-6361:20053555

\bibitem[Granato, \& Danese(1994)]{Granato94} Granato, G.~L., \& Danese, L.\ 1994, \mnras, 268, 235

\bibitem[Gratadour et al.(2015)]{Gratadour15} Gratadour, D., Rouan, D., Grosset, L., et al.\ 2015, \aap, 581, L8. doi:10.1051/0004-6361/201526554

\bibitem[Gravity Collaboration et al.(2020)]{Gravity20} Gravity Collaboration, Pfuhl, O., Davies, R., et al.\ 2020, \aap, 634, A1. doi:10.1051/0004-6361/201936255

\bibitem[Greenhill \& Gwinn(1997)]{Greenhill97} Greenhill, L.~J. \& Gwinn, C.~R.\ 1997, \apss, 248, 261. doi:10.1023/A:1000554317683

\bibitem[Harris et al.(2020)]{Harris20} Harris, C.~R., Millman, K.~J., van der Walt, S.~J., et al.\ 2020, \nat, 585, 357. doi:10.1038/s41586-020-2649-2


\bibitem[H{\"o}nig \& Kishimoto(2010)]{Hoenig10B} H{\"o}nig, S.~F. \& Kishimoto, M.\ 2010, \aap, 523, A27. doi:10.1051/0004-6361/200912676

\bibitem[H{\"o}nig et al.(2013)]{Hoenig13} H{\"o}nig, S.~F., Kishimoto, M., Tristram, K.~R.~W., et al.\ 2013, \apj, 771, 87. doi:10.1088/0004-637X/771/2/87

\bibitem[H{\"o}nig \& Kishimoto(2017)]{Hoenig17} H{\"o}nig, S.~F., \& Kishimoto, M.\ 2017, \apjl, 838, L20

\bibitem[Hunter(2007)]{Hunter07} Hunter, J.~D.\ 2007, Computing in Science and Engineering, 9, 90. doi:10.1109/MCSE.2007.55

\bibitem[Imanishi et al.(2018)]{Imanishi18} Imanishi, M., Nakanishi, K., Izumi, T., et al.\ 2018, \apjl, 853, L25. doi:10.3847/2041-8213/aaa8df

\bibitem[Impellizzeri et al.(2019)]{Impellizzeri19} Impellizzeri, C.~M.~V., Gallimore, J.~F., Baum, S.~A., et al.\ 2019, \apjl, 884, L28. doi:10.3847/2041-8213/ab3c64

\bibitem[Jaffe et al.(2004)]{Jaffe04} Jaffe, W., Meisenheimer, K., R{\"o}ttgering, H.~J.~A., et al.\ 2004, \nat, 429, 47

\bibitem[Kishimoto et al.(2007)]{Kishimoto07} Kishimoto, M., H{\"o}nig, S.~F., Beckert, T., et al.\ 2007, \aap, 476, 713

\bibitem[Khouri et al.(2020)]{Khouri20} Khouri, T., Vlemmings, W.~H.~T., Paladini, C., et al.\ 2020, \aap, 635, A200.

\bibitem[Kobayashi et al.(2011)]{Kobayashi11} Kobayashi, H., Kimura, H., Watanabe, S.-. i ., et al.\ 2011, Earth, Planets, and Space, 63, 1067. doi:10.5047/eps.2011.03.012

\bibitem[Li \& Draine(2001)]{Li-Draine01} Li, A. \& Draine, B.~T.\ 2001, \apj, 554, 778. doi:10.1086/323147

\bibitem[Lira et al.(2013)]{Lira2013} Lira, P., Videla, L., Wu, Y., et al.\ 2013, \apj, 764, 159

\bibitem[Liu et al.(2016)]{Liu16} Liu, J., Liu, Y., Li, X., et al.\ 2016, \mnras, 459, L100. doi:10.1093/mnrasl/slw042

\bibitem[Liu et al.(2019)]{Liu19} Liu, J., H{\"o}nig, S.~F., Ricci, C., et al.\ 2019, \mnras, 490, 4344. doi:10.1093/mnras/stz2908

\bibitem[L{\'o}pez-Gonzaga et al.(2014)]{Lopez-Gonzaga14} L{\'o}pez-Gonzaga, N., Jaffe, W., Burtscher, L., et al.\ 2014, \aap, 565, A71. doi:10.1051/0004-6361/201323002

\bibitem[Lopez-Rodriguez et al.(2016)]{Lopez-Rodriguez16} Lopez-Rodriguez, E., Packham, C., Roche, P.~F., et al.\ 2016, \mnras, 458, 3851. doi:10.1093/mnras/stw541

\bibitem[Lopez-Rodriguez et al.(2018)]{Lopez-Rodriguez18} Lopez-Rodriguez, E., Fuller, L., Alonso-Herrero, A., et al.\ 2018, arXiv:1804.04134 

\bibitem[Lopez-Rodriguez et al.(2020)]{Lopez-Rodriguez20} Lopez-Rodriguez, E., Alonso-Herrero, A., Garc{\'\i}a-Burillo, S., et al.\ 2020, \apj, 893, 33. doi:10.3847/1538-4357/ab8013

\bibitem[Lyu et al.(2014)]{Lyu14} Lyu, J., Hao, L., \& Li, A.\ 2014, \apjl, 792, L9. doi:10.1088/2041-8205/792/1/L9

\bibitem[Lyu \& Rieke(2021)]{Lyu21} Lyu, J. \& Rieke, G.~H.\ 2021, \apj, 912, 126. doi:10.3847/1538-4357/abee14

\bibitem[Marco \& Alloin(2000)]{Marco2000} Marco, O., \& Alloin, D.\ 2000, \aap, 353, 465

\bibitem[Mart{\'\i}nez-Paredes et al.(2020)]{Martinez-Paredes20} Mart{\'\i}nez-Paredes, M., Gonz{\'a}lez-Mart{\'\i}n, O., Esparza-Arredondo, D., et al.\ 2020, \apj, 890, 152. doi:10.3847/1538-4357/ab6732

\bibitem[Mason et al.(2006)]{Mason06} Mason, R.~E., Geballe, T.~R., Packham, C., et al.\ 2006, \apj, 640, 612. doi:10.1086/500299

\bibitem[Mathis et al.(1977)]{Mathis77} Mathis, J.~S., Rumpl, W., \& Nordsieck, K.~H.\ 1977, \apj, 217, 425. doi:10.1086/155591

\bibitem[Matter et al.(2020)]{Matter20} Matter, A., Pignatale, F.~C., \& Lopez, B.\ 2020, \mnras, 497, 2540.

\bibitem[Miller \& Antonucci(1983)]{Miller83} Miller, J.~S. \& Antonucci, R.~R.~J.\ 1983, \apjl, 271, L7. doi:10.1086/184082

\bibitem[Min et al.(2003)]{Min03} Min, M., Hovenier, J.~W., \& de Koter, A.\ 2003, \aap, 404, 35. doi:10.1051/0004-6361:20030456

\bibitem[Min et al.(2006)]{Min06} Min, M., Dominik, C., Hovenier, J.~W., et al.\ 2006, \aap, 445, 1005. doi:10.1051/0004-6361:20053212

\bibitem[Min et al.(2007)]{Min07} Min, M., Waters, L.~B.~F.~M., de Koter, A., et al.\ 2007, \aap, 462, 667

\bibitem[Nasa High Energy Astrophysics Science Archive Research Center (Heasarc)(2014)]{Heasarc14} Nasa High Energy Astrophysics Science Archive Research Center (Heasarc)\ 2014, Astrophysics Source Code Library. ascl:1408.004

\bibitem[Nenkova et al.(2008a)]{Nenkova08A} Nenkova, M., Sirocky, M.~M., Ivezi{\'c}, {\v Z}., \& Elitzur, M.\ 2008, \apj, 685, 147-159

\bibitem[Nenkova et al.(2008b)]{Nenkova08B} Nenkova, M., Sirocky, M.~M., Nikutta, R., Ivezi{\'c}, {\v Z}., \& Elitzur, M.\ 2008, \apj, 685, 160-180

\bibitem[Netzer(1987)]{Netzer87} Netzer, H.\ 1987, \mnras, 225, 55

\bibitem[Netzer(2013)]{Netzer13} Netzer, H.\ 2013, The Physics and Evolution of Active Galactic Nuclei, by Hagai Netzer, Cambridge, UK: Cambridge University Press, 2013

\bibitem[Nikutta et al.(2009)]{Nikuta09} Nikutta, R., Elitzur, M., \& Lacy, M.\ 2009, \apj, 707, 1550. doi:10.1088/0004-637X/707/2/1550

\bibitem[Oliphant(2007)]{Oliphant07} Oliphant, T.~E.\ 2007, Computing in Science and Engineering, 9, 10. doi:10.1109/MCSE.2007.58

\bibitem[Olofsson et al.(2009)]{Olofsson09} Olofsson, J., Augereau, J.-C., van Dishoeck, E.~F., et al.\ 2009, \aap, 507, 327.

\bibitem[Pasetto et al.(2019)]{Pasetto19} Pasetto, A., Gonz{\'a}lez-Mart{\'\i}n, O., Esparza-Arredondo, D., et al.\ 2019, The Astrophysical Journal, 872, 69

\bibitem[P{\'e}rez et al.(2012)]{Perez12} P{\'e}rez, L.~M., Carpenter, J.~M., Chandler, C.~J., et al.\ 2012, \apjl, 760, L17.

\bibitem[Pier, \& Krolik(1992)]{Pier92} Pier, E.~A., \& Krolik, J.~H.\ 1992, \apj, 401, 99

\bibitem[Pinte et al.(2009)]{Pinte09} Pinte, C., Harries, T.~J., Min, M., et al.\ 2009, \aap, 498, 967. doi:10.1051/0004-6361/200811555


\bibitem[Raban et al.(2009)]{Raban09} Raban, D., Jaffe, W., R{\"o}ttgering, H., et al.\ 2009, \mnras, 394, 1325. doi:10.1111/j.1365-2966.2009.14439.x

\bibitem[Ramos Almeida et al.(2009)]{Ramos-Almeida09} Ramos Almeida, C., Levenson, N.~A., Rodr{\'\i}guez Espinosa, J.~M., et al.\ 2009, \apj, 702, 1127. doi:10.1088/0004-637X/702/2/1127

\bibitem[Ramos Almeida et al.(2011)]{Ramos-Almeida11} Ramos Almeida, C., Levenson, N.~A., Alonso-Herrero, A., et al.\ 2011, \apj, 731, 92

\bibitem[Ramos Almeida et al.(2014)]{Ramos-Almeida14} Ramos Almeida, C., Alonso-Herrero, A., Levenson, N.~A., et al.\ 2014, \mnras, 439, 3847. doi:10.1093/mnras/stu235

\bibitem[Ramos Almeida \& Ricci(2017)]{Ramos-Almeida17} Ramos Almeida, C. \& Ricci, C.\ 2017, Nature Astronomy, 1, 679. doi:10.1038/s41550-017-0232-z

\bibitem[Rojas Lobos et al.(2018)]{Rojas18} Rojas Lobos, P.~A., Goosmann, R.~W., Marin, F., et al.\ 2018, \aap, 611, A39. doi:10.1051/0004-6361/201731331

\bibitem[Schartmann et al.(2005)]{Schartmann05} Schartmann, M., Meisenheimer, K., Camenzind, M., et al.\ 2005, \aap, 437, 861

\bibitem[Shao et al.(2017)]{Shao17} Shao, Z., Jiang, B.~W., \& Li, A.\ 2017, \apj, 840, 27. doi:10.3847/1538-4357/aa6ba4


\bibitem[Siebenmorgen et al.(2015)]{Siebenmorgen15} Siebenmorgen, R., Heymann, F., \& Efstathiou, A.\ 2015, \aap, 583, A120

\bibitem[Stalevski et al.(2012)]{Stalevski12} Stalevski, M., Fritz, J., Baes, M., et al.\ 2012, \mnras, 420, 2756

\bibitem[Stalevski et al.(2016)]{Stalevski16} Stalevski, M., Ricci, C., Ueda, Y., et al.\ 2016, Monthly Notices of the Royal Astronomical Society, 458, 2288

\bibitem[Stalevski et al.(2019)]{Stalevski19} Stalevski, M., Tristram, K.~R.~W., \& Asmus, D.\ 2019, \mnras, 484, 3334. doi:10.1093/mnras/stz220

\bibitem[Testi et al.(2001)]{Testi01} Testi, L., Natta, A., Shepherd, D.~S., et al.\ 2001, \apj, 554, 1087. doi:10.1086/321406

\bibitem[Tomono et al.(2001)]{Tomono01} Tomono, D., Doi, Y., Usuda, T., et al.\ 2001, \apj, 557, 637. doi:10.1086/322262


\bibitem[Urry \& Padovani(1995)]{Urry-Padovani95} Urry, C.~M. \& Padovani, P.\ 1995, \pasp, 107, 803. doi:10.1086/133630

\bibitem[van Bemmel, \& Dullemond(2003)]{Van_Bemmel03} van Bemmel, I.~M., \& Dullemond, C.~P.\ 2003, \aap, 404, 1

\bibitem[Venanzi et al.(2020)]{Venanzi20} Venanzi, M., H{\"o}nig, S., \& Williamson, D.\ 2020, \apj, 900, 174. doi:10.3847/1538-4357/aba89f

\bibitem[Virtanen et al.(2020)]{Virtanen20} Virtanen, P., Gommers, R., Oliphant, T.~E., et al.\ 2020, Nature Methods, 17, 261. doi:10.1038/s41592-019-0686-2


\bibitem[Wada(2012)]{Wada12} Wada, K.\ 2012, \apj, 758, 66. doi:10.1088/0004-637X/758/1/66

\bibitem[Wang et al.(2020)]{Wang20} Wang, J.-M., Songsheng, Y.-Y., Li, Y.-R., et al.\ 2020, \mnras, 497, 1020. doi:10.1093/mnras/staa1985

\bibitem[Weingartner \& Draine(2001)]{Weingartner01} Weingartner, J.~C. \& Draine, B.~T.\ 2001, \apj, 548, 296. doi:10.1086/318651

\bibitem[Wittkowski et al.(2004)]{Wittkowski04} Wittkowski, M., Kervella, P., Arsenault, R., et al.\ 2004, \aap, 418, L39. doi:10.1051/0004-6361:20040118



\end{thebibliography}
\end{document}